\documentclass[12pt]{article}
\usepackage{graphicx}
\begin{document}
\rightline{NKU-2011-SF1}
\bigskip
\begin{center}
{\Large\bf Null Geodesics of Charged Black Holes in String Theory}

\end{center}
\hspace{0.4cm}
\begin{center}
Sharmanthie Fernando \footnote{fernando@nku.edu}\\ 
{\small\it Department of Physics \& Geology}\\
{\small\it Northern Kentucky University}\\
{\small\it Highland Heights}\\
{\small\it Kentucky 41099}\\
{\small\it U.S.A.}\\

\end{center}
\begin{center}
{\bf Abstract}
\end{center}

\hspace{0.7cm} 
In this paper, we investigate the null geodesics of the static charged black hole in heterotic string theory. A detailed analysis of the geodesics are done in the Einstein frame as well as in the string frame. In the Einstein frame, the geodesics are solved exactly in terms of the Jacobi-elliptic integrals for all possible energy levels and angular momentum of the photons. In the string frame, the geodesics are presented for the circular orbits. As a physical application of the null geodesics, we have obtained the angle of deflection for the photons  and the quasinormal modes of a massless scalar field in the eikonal limit.


\section{Introduction }

String theory has become the leading candidate to unify gravity with the rest of the fundamental forces in nature. Therefore, studies of black holes in string theory takes an important place in current research. In this paper, we wish to study four dimensional spherically symmetric static charged black holes in heterotic string theory. The action of the low-energy   heterotic string theory in four dimensions is given as follows,
\begin{equation}
S= \frac{1}{ 16 \pi} \int d^4 x\sqrt{- det~g}  \left(R - \frac{1}{12} e^{ - 4 \Phi} H_{\mu\nu\rho}H^{\mu\nu\rho} - 2 (\bigtriangledown \Phi )^2 - e^{ - 2 \Phi} F_{\mu\nu}F^{\mu\nu}\right)
\end{equation}
Here $g_{\mu\nu}$ is the metric, 
$F_{\mu\nu}=\partial_\mu A_\nu-\partial_\nu A_\mu$ is the field strength
corresponding to the Maxwell field $A_\mu$, $\Phi$ is the dilaton field,
and,
\begin{equation}
H_{\mu\nu\rho} =\partial_\mu B_{\nu\rho} +{~\rm cyclic~permutations}
-(\Omega_3(A))_{\mu\nu\rho}
\end{equation}
where $B_{\mu\nu}$ is an antisymmetric tensor gauge field. The gauge Chern-Simons term is given as,
\begin{equation}
(\Omega_3(A))_{\mu\nu\rho}={1\over 4} (A_\mu F_{\nu\rho}+ {~\rm cyclic
{}~permutations})
\end{equation}
For more details, see\cite{sen1}\cite{sen2}.

In this paper, we will focus on the solution to the  action in eq.(1) with the fields $H_{ \mu \nu \alpha}$ and $ B_{ \mu \nu}$ set to zero. Hence the heterotic string action simplifies to,
\begin{equation}
S = \frac{1}{ 16 \pi} \int d^4x \sqrt{-g} \left[ R - 2 (\bigtriangledown \Phi )^2 -
e^{-2  \Phi} F_{\mu \nu} F^{\mu \nu}  \right]
\end{equation}
Here $\Phi$ is the dilaton field, $R$ is the scalar curvature and $F_{\mu \nu}$ is the Maxwell's field strength. The static charged black hole solutions to the above action were  found first by Gibbons and Maeda\cite{gib}. It was also independently found  by Garfinkle, Horowitz and Strominger \cite{gar} few years later. In the rest of the paper, we will refer to this black hole as the Gibbons-Maeda-Garfinkle-Horowitz-Strominger (GMGHS) black hole.

The main objective of this paper is to study the geodesic structure of massless particles of the  GMGHS black hole. Studies of test particles, both massive and massless is one way to understand the gravitational field around a black hole. Since the motion of test particles around black holes are determined by the geodesic structure, a detailed analysis of geodesics are of great significance. Theoretical predictions such as light deflection, gravitational time-delay,  perihelion shift and Lense-Thirring effect etc are physical elements that can be compared with observations when one study black holes in the universe. All of such phenomenon are related to the geodesics of black holes. Furthermore, the study of orbits of test particles is of astrophysical relevance when it comes to flow of particles in accretion disks around black holes. 

Geodesics of black holes are studied extensively in the literature. There are many works related to the geodesics of  well known charged black hole in general relativity which is the Reissner-Nordstr$\ddot{o}$m black hole. Pugliese et.al.\cite{pug1}\cite{pug2} have studied the geodesics of neutral as well as charged test particles of the Reissner-Nordstr$\ddot{o}$m black hole recently. Analytical solutions of the electrically and magnetically charged test particles were discussed by Grunau and Kagramanova\cite{gru}. Chandrasekhar\cite{chandra} and Hackmann et.al.\cite{hack2} also have  studied the geodesics of the Reissner-Nordstr$\ddot{o}$m black hole.

There are few works that we like to mention that have addressed geodesics related to string black holes. Kinematics of the time-like geodesic congruences of string black holes in two and four dimensions were discussed by Dasgupta et.al in\cite{das}. Motion of test particles around a charged dilatonic black hole is discussed by Maki and Shiraishi \cite{maki}. Hidden symmetries, null geodesics and photon capture of the Sen black hole, which is the rotating version of the GMGHS black hole, is discussed by Hioki and Miyamoto\cite{hioki}. There, comments of the circular photon orbits of the GMGHS black hole are given. Bounded radial geodesics of the Sen black hole is discussed by Blaga and Blaga\cite{blaga}. Gravitational lensing by a charged black hole in string theory was studied by Bhadra\cite{bhadra}. The geodesics of the 2+1 dimensional  string black hole was studied by Fernando et.al\cite{fer1}.

The paper is organized as follows. In section 2, we will present the GMGHS black hole in the Einstein frame and string frame. A comparison with the Reissner-Nordstr$\ddot{o}$m black hole is included in this section. The geodesic equations for the GMGHS black hole in the Einstein frame is derived  in section 3. A detailed discussion of the null geodesics in the Einstein frame follows in section 4. In section 5, bending of light and quasinormal modes are discussed as an application of the null geodesics. In section 6, the geodesic equations in string frame is presented. Null geodesics in string frame is presented in section 7. Finally, the conclusion is given in section 8.

\section{ The GMGHS charged black hole in 
string theory}

\subsection{ GMGHS black hole in Einstein frame}

The GMGHS black hole solution to the action in eq.(4) is given by,
\begin{equation}
ds^2_E = - \left(1-\frac{2M}{r}\right) dt^2 + \frac{1}{\left(1-\frac{2M}{r}\right)} dr^2
+ r\left(r-\frac{Q^2} {M} \right) ( d \theta^2 + sin^2(\theta) d \phi^2)
\end{equation}
Here, the electric field strength and the dilaton field are given by,
\begin{equation}
{ F_{rt}=\frac{Q}{r^2};\qquad e^{2\Phi}
= 1-\frac{Q^2}{Mr}}
\end{equation}
In the $r-t$ plane, the metric simplifies to,
\begin{equation}
ds^2_E = - \left(1-\frac{2M}{r}\right) dt^2 + \frac{1}{\left(1-\frac{2M}{r}\right)} dr^2
\end{equation}
Hence, the metric is identical to the Schwarzschild black hole metric. There is an event horizon at $ r = 2 M$. How ever, the area of the sphere of the string black hole is smaller and the area approaches  zero when $ r = Q^2/M$. Therefore, $ r = Q^2/M$ surface is singular. For $Q^2 \leq 2M^2$, the singular surface is  inside the event horizon and the Penrose diagram is identical to the Schwarzschild black hole. See the review on black holes by Townsend\cite{town} for more details about the Schwarzschild black hole. When $ Q^2=2 M^2$, the singular surface  coincide with the horizon. This is the extremal limit where a transition between the black hole and the naked singularity occurs.

\subsection{ GMGHS black hole in string frame}

In the above presentation of the GMGHS black hole, we have written it in the so-called ``Einstein frame''. In Einstein frame, the action is the Einstein-Hilbert action that is used in general relativity. However, when dealing with low-energy string theory and it's solutions, there is another frame called ``string frame'' in which the metric can be written. String frame is some times preferred because it is the metric that strings directly coupled to.  An interesting discussion on the relations between the string frame and the Einstein frame is given by Casadio and Harms\cite{harms}. There are differences in some of the physical properties of  the black holes in two frames. Casadio and Harms have  done a detailed analysis of two black holes in dilaton gravity comparing the properties of the black holes in the string frame(SF) vs Einstein frame(EF). It is given that the area  of the black hole outer horizon is different from SF to EF.  Hence they conclude that the emission rate is higher in SF compared to EF. In the rotating dilaton black hole, computations are presented to show that the gyromagnetic ratio $g$ is different in two frames. Also the scalar curvature $R$ is different in two frames. Which frame is more suitable to describe the current state of the universe is an open question which needs to be settled by experiments. \\

The string metric is given by,
\begin{equation}
ds^2_{string} = - \frac{\left(1-\frac{2m}{\hat{r}}\right)}{ \left( 1 + \frac{ 2 m \sinh^2(\alpha)}{\hat{r}} \right)^2} dt^2 + \frac{d\hat{r}^2}{\left(1-\frac{2m}{\hat{r}}\right)} 
+ \hat{r}^2 ( d \theta^2 + sin^2(\theta) d \phi^2)
\end{equation}
The dilaton field in the string metric is given by,
\begin{equation}
e^{ - 2\Phi_{string}} = 1 + \frac{2 m sinh^2 \alpha}{\hat{r}}
\end{equation}
The physical mass M and the charge Q are related to $m$ and $\alpha$  as,
\begin{equation}
M= m  \cosh^2(\alpha),~~~~ Q=  \sqrt {2}m \sinh(\alpha) \cosh(\alpha)
\end{equation}
The metric in two frames are related by a coordinate transformation and a conformal transformation as follows:
First perform a conformal transformation on the string metric as,
\begin{equation}
g^E_{\mu\nu} = e^{-2\Phi_{string}} g^S_{\mu\nu}
\end{equation}
and then, do a coordinate transformation as,
\begin{equation}
\hat{r} = r - 2 m sinh^2 \alpha
\end{equation}
These two transformations will lead to the GMGHS metric in the Einstein frame.

Note that the horizon in the string frame is
\begin{equation}
\hat{r}_h^{string} =  2 m = \frac{ 2 M}{   \cosh^2(\alpha)}
\end{equation}
Since $cosh(\alpha) \geq 1$, the horizon in the string frame is smaller than the one in the Einstein frame with the same mass. Therefore, the black hole seems ``smaller'' in string frame.

\subsection{ Comparison of the GMGHS black hole with the Reissner-Nordstr$\ddot{o}$m black hole}

The well known charged black hole in Einstein-Maxwell gravity is given by,
\begin{equation}
ds^2_E = - \left(1-\frac{2M}{r} + \frac{ Q^2}{r^2}\right) dt^2 + \frac{1}{\left(1-\frac{2M}{r} + \frac{Q^2}{r^2} \right)} dr^2
+ r^2  d \Omega^2
\end{equation}
This metric has mass $M$ and charge $Q$ and have two horizons,
\begin{equation}
r_{\pm} = M \pm \sqrt{ M^2 - Q^2}
\end{equation}
Reissner-Nordstr$\ddot{o}$m black hole become a naked singularity when $Q^2 = M^2$.

In comparison with the GMGHS black hole, there are similarities as well as differences. Both represents black holes for small $Q/M$ and become naked singularities for large $Q/M$. The GMGHS black hole lack inner horizon while the Reissner-Nordstr$\ddot{o}$m black hole does. For both solutions,  maximal value of charge exits to separate black hole from the naked singularity. How ever, the extrem solutions do have different properties\cite{gar}. More details comparing the GMGHS black hole to Riessner-Nordstr$\ddot{o}$m black hole and to the Schwarzschild black hole can be found in the paper by Garfinkle et.al.\cite{gar} and the review by Horowitz\cite{hor}.

\section{ Geodesics in Einstein Frame}

First we will derive the geodesic equations for neutral particles around the GMGHS black hole. We will follow the same approach given in the well known book by Chandrasekhar\cite{chandra}. The metric is written as,
\begin{equation}
ds^2 = -f(r) dt^2 + f(r)^{-1} dr^2 +  R(r)^2 ( d \theta^2 + sin^2(\theta) d \phi^2)
\end{equation}
Here,
\begin{equation}
f(r) = \left(1-\frac{2M}{r}\right)
\end{equation}
\begin{equation}
R(r)^2 = r\left(r-\frac{Q^2} {M} \right) = r ( r -a)
\end{equation}
and, 
\begin{equation}
 a = \frac{Q^2}{M}
\end{equation}
Equations governing the geodesics in this space-time can be derived from the Lagrangian equation,
\begin{equation}
{\cal{L}} =  - \frac{1}{2} \left( - f(r) \left( \frac{dt}{d\tau} \right)^2 +  \frac{1}{f(r)}\left( \frac{dr}{d \tau} \right)^2 + R(r)^2 \left(\frac{d \theta}{d \tau} \right)^2 + R(r)^2 sin^2 \theta \left( \frac{d \phi }{d \tau} \right)^2 \right)
\end{equation}
Here, $\tau$ is an affine parameter along the geodesics. The canonical mometa corresponding to each coordinate is given as,
\begin{equation}
p_t = \frac{d {\cal{L}} }{d \dot{t}} = f \dot{t}
\end{equation}
\begin{equation}
p_{r} = - \frac{d {\cal{L}} }{d \dot{r}} = \frac{ \dot{r}}{f}
\end{equation}
\begin{equation}
p_{\theta}  = - \frac{d {\cal{L}} }{d \dot{\theta}} =  R(r)^2 \dot{\theta}
\end{equation}
\begin{equation}
p_{\phi}  = - \frac{d {\cal{L}} }{d \dot{\phi}} =  R(r)^2 sin^2 \theta \dot{\phi}
\end{equation}
Since the GMGHS black hole have  two Killing vectors $\partial_t$ and $\partial_{\phi}$, there are two constants of motion which can be labeled as $E$ and $L$ given as,
\begin{equation}
p_t = f \dot{t} = E
\end{equation}
\begin{equation}
p_{\phi} =    R(r)^2 sin^2 \theta \dot{\phi} = L
\end{equation}
Furthermore, from the Euler-Lagrangian  equation of motion,
\begin{equation}
\frac{d}{d \tau} \left(\frac{d {\cal{L}} }{d \dot{\theta}} \right) - \frac{d {\cal{L}} }{d \theta} =0 \Rightarrow -\frac{d p_{\theta}}{d \tau}  - \frac{d {\cal{L}} }{d \theta}=0
\Rightarrow 
\frac{d (- R(r)^2 \dot{\theta}) }{d \tau} + R(r)^2 sin \theta  cos \theta  \left( \frac{d \phi }{d \tau}\right)^2=0
\end{equation}
Hence, if we choose $\theta = \pi/2$ and $\dot{\theta} = 0$ as the initial conditions,  the eq.(25) leads to,
\begin{equation}
\frac{d (R(r)^2 \dot{\theta}) }{d \tau} =  R(r)^2 \ddot{\theta} +  \frac{d R(r)^2 }{d \tau} \dot{ \theta} = 0
\end{equation}
Therefore, $\ddot{\theta} =0$. $\theta$ will remain at $\pi/2$ and the geodesics will be described in an invariant plane at $ \theta = \pi/2$.
From eq.(25) and eq.(26),
\begin{equation}
R(r)^2 \dot{\phi} = L; \hspace{1.0cm} f(r) \dot{t} = E
\end{equation}
With $\dot{t}$ and $\dot{\phi}$ given by equation(29), the Lagrangian in eq.(20) simplifies to be,
\begin{equation}
\dot{r}^2 + f(r) \left(  \frac{L^2}{R(r)^2} + h   \right)  = E^2
\end{equation}
Here, $2 {\cal{L}}=h$. $h=1$ corresponds to time-like geodesics and $h=0$ corresponds to null geodesics. For a time-like geodesic, $\tau$ may be identified with proper time of the particle describing the geodesic. Comparing eq.(28) with $\dot{r}^2 + V_{eff}= E^2$, one get the effective potential, which depend on $E$ and $L$ as follows:
\begin{equation}
V_{eff} = \left(  \frac{L^2}{R(r)^2} + h  \right) f(r) 
\end{equation}
By eliminating the parameter $\tau$ from the equations(29) and (30), one can get a relation between $\phi$ and $r$ as follows;
\begin{equation}
\frac{ d \phi} {d r} =  \frac{L}{R(r)^2} \frac{1}{ \sqrt{  ( E^2 - V_{eff})} }
\end{equation}

\section{Null Geodesics in Einstein frame}

In the above formalism, $h=0$ for null geodesics.

\subsection{ Radial null geodesics}

The radial geodesics corresponds to the motion of particles with zero angular momentume ($L =0$). Hence for radial null geodesics, the effective potential,
\begin{equation}
V_{eff} = 0
\end{equation}
The two equations for $\dot{t}$ and $\dot{r}$ simplifies to,
\begin{equation}
\dot{ r} = \pm E; \hspace{1 cm} \dot{t} = \frac{E}{f(r)}
\end{equation}
The above two equations lead to,
\begin{equation}
\frac{ dt}{ dr} = \pm\frac{1}{f(r) } = \pm \frac{ 1} { (1 - \frac{ 2 M}{ r}) }
\end{equation}
The above equation can be integrated to give the coordinate time $t$ as a function of $r$ as,
\begin{equation}
t = \pm \left( r + 2 M ln \left( \frac{r}{ 2M} -1 \right) \right) + const_{\pm}
\end{equation}
When $ r \rightarrow 2 M$, $ t \rightarrow \infty$. On the other hand, the proper time can be obtained by integrating,
\begin{equation}
\frac{ d \tau}{ dr} = \pm\frac{1}{ E } 
\end{equation}
which leads to,
\begin{equation}
 \tau = \pm \frac{ r}{ E} + const_{\pm}
\end{equation}
When $ r \rightarrow 2 M$, $ \tau \rightarrow \pm  \frac{ 2 M} { E} $, which is finite. Hence the proper time is finite while the coordinate time is infinite. This is the same results one would obtain for the Schwarzschild black hole\cite{chandra}.

\subsection{ Geodesics with angular momentum ($ L \neq 0$)}

In this section, we will study the null geodesics with angular momentum.

\subsubsection{ Effective potential}

In this case, 
\begin{equation}
 V_{eff} = f(r) \frac{L^2}{ R(r)^2}
\end{equation}
For $ r = 2 M$, $ V_{eff} = 0$ and for $ r \rightarrow \infty$, $ V_{eff} \rightarrow 0$. In the Fig. 1, the  $V_{eff}$ is given for various values of $a$. One can note that the height is higher for the string black hole in comparison to the Schwarzschild black hole.

\begin{center}
\scalebox{.9}{\includegraphics{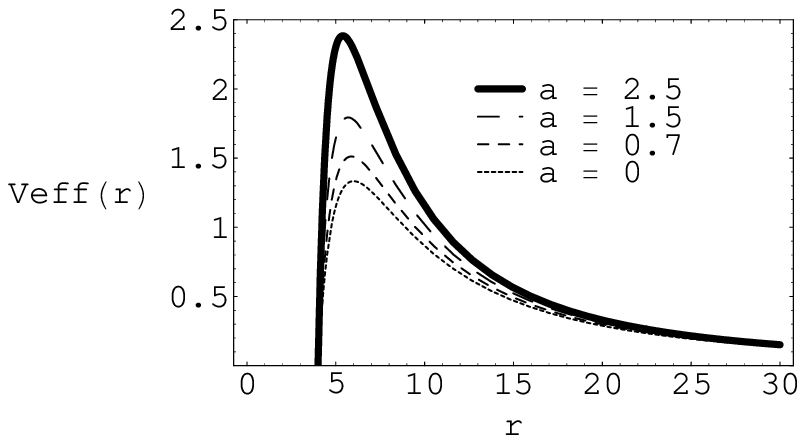}}\\
\vspace{0.1cm}
\end{center}
Figure 1. The graph shows the relation of $V_{eff}$ with the parameter $a$. Here, $M =2, L =12$\\

In Fig.2  the effective potential for various values of the angular momentum $L$ is given. As expected, the effective potential is larger for large $L$.\\
\begin{center}
\scalebox{.9}{\includegraphics{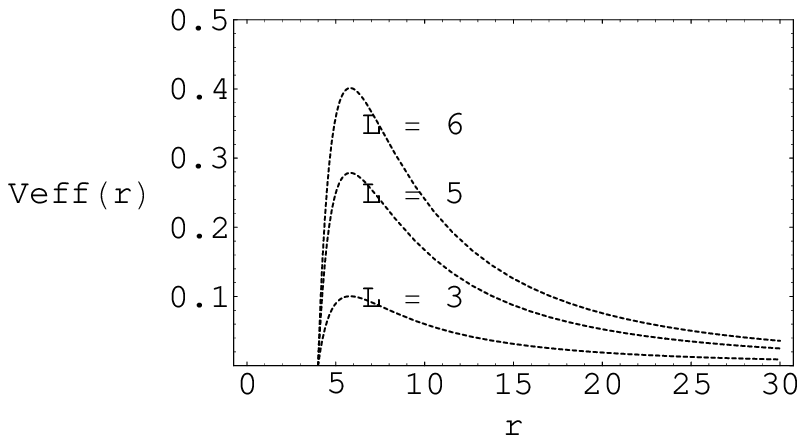}}\\
\vspace{0.1cm}
\end{center}
Figure 2. The graph shows the relation of $V_{eff}$ with the angular momentum $L$. Here $ M = 2 , a = 1.5$\\

Since $\dot{r}^2 + V_{eff}= E^2$,  the motion of the particles depend on the energy levels. In the Fig.3, the effective potential is plotted and three energy levels, $E_1$, $E_c$ and $E_2$ are given which corresponds to three different scenarios of motion of the particles which is described below.\\

{\bf Case 1:}  { $E = E_c$}

Here, $ E^2 - V_{eff} = 0$ and $\dot{r} =0$ leading to circular orbits. However, due to the nature of the potential at $ r=r_c$, these are unstable circular orbits which will be presented in detail in the  section(4.2.3).\\

{\bf Case 2:}  { $E = E_2$}

Here, $ E^2 - V_{eff} \geq 0$ only for $ r \leq r_A$ and $ r \geq r_P$ ( as indicated in Fig.3). Hence the motion is possible only in those regions. If the photons start at infinity, it will fall to a minimum radius $r_P$ and flies back to infinity. Hence the photons are deflected. If the photons start at $ r = r_A$, they will fall into the singularity crossing the horizon at $ r = 2M$.\\

{\bf Case 3:}  { $E = E_1$}

Here, $ E^2 - V_{eff} \geq 0$ and $\dot{r} >0$ for all $r$ values. Hence the photons coming from infinity cross the horizon at $ r = 2m$ and falls into the singularity at $ r =a$.

\begin{center}
\scalebox{.9}{\includegraphics{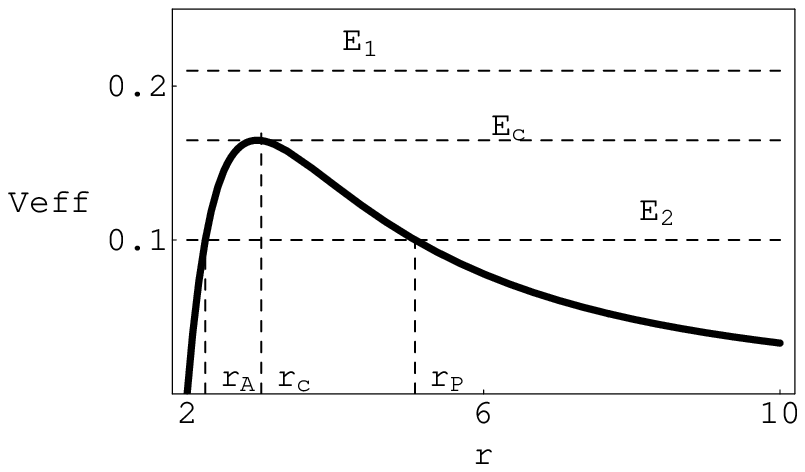}}\\
\vspace{0.2cm}
\end{center}
Figure 3. The graph shows the relation of $V_{eff}$ with the energy $E$. Here, $ M =1, a = 0.3$ and $L =2$.

\subsubsection{Analysis of the geodesics with the variable $ u = \frac{1}{r}$}

One can also study the orbits by doing a well known change of variable $ u = \frac{1}{r}$. Then the eq.(30) can be rewritten in terms of $u$ and $\phi$ as,
\begin{equation}
\left(\frac{ d u}{ d \phi} \right)^2 = f(u)
\end{equation}
where,
\begin{equation}
f(u) = - 2 a M u^4 + ( a + 2 M) u^3 +  u^2 \left( a^2 \frac{ E^2}{L^2} -1 \right) - 2 a \frac{E^2}{L^2} u + \frac{ E^2}{L^2}
\end{equation}
The function $f(u)$ for general values of $ M, a, E$ and $L$ is given in the Fig.4. 

\begin{center}
\scalebox{0.9}{\includegraphics{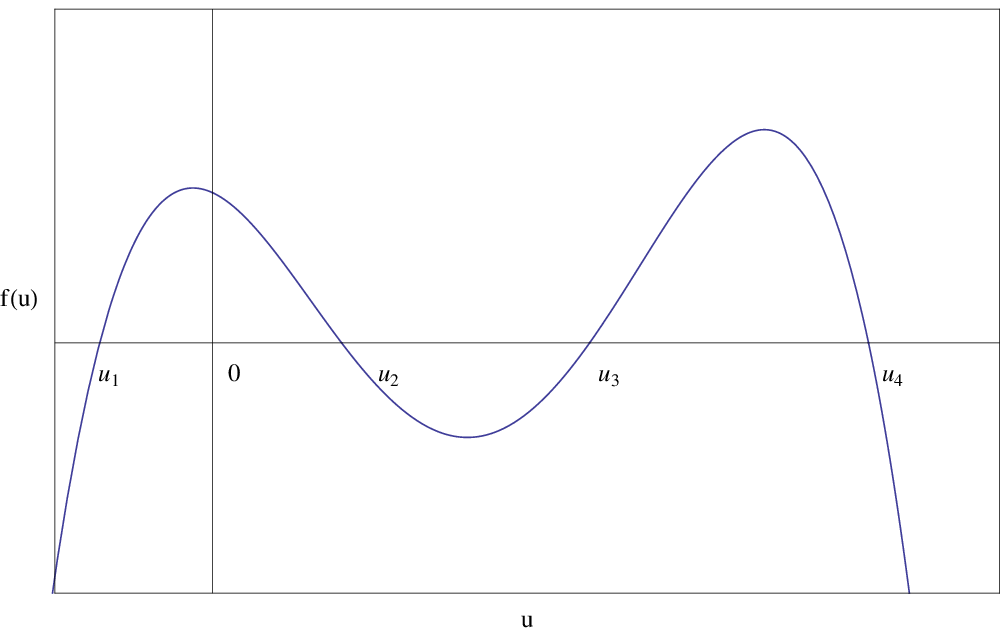}}\\
\vspace{0.2cm}
\end{center}
Figure 4. The graph shows the function $f(u)$ for $ M =0.5, a = 0.6, L = 30$ and $ E = 9$.\\

When $ a \rightarrow 0$, $ f(u) \rightarrow 2 M u^3 - u^2 + \frac{E^2}{L^2}$ as expected for the Schwarzschild black hole \cite{chandra}. 
When $ a \rightarrow 0$, $f(u)$ has maximum three real roots as described in the book by Chandrasekhar\cite{chandra}. The function for $a =0$ is given in the Fig.5.

\begin{center}
\scalebox{.9}{\includegraphics{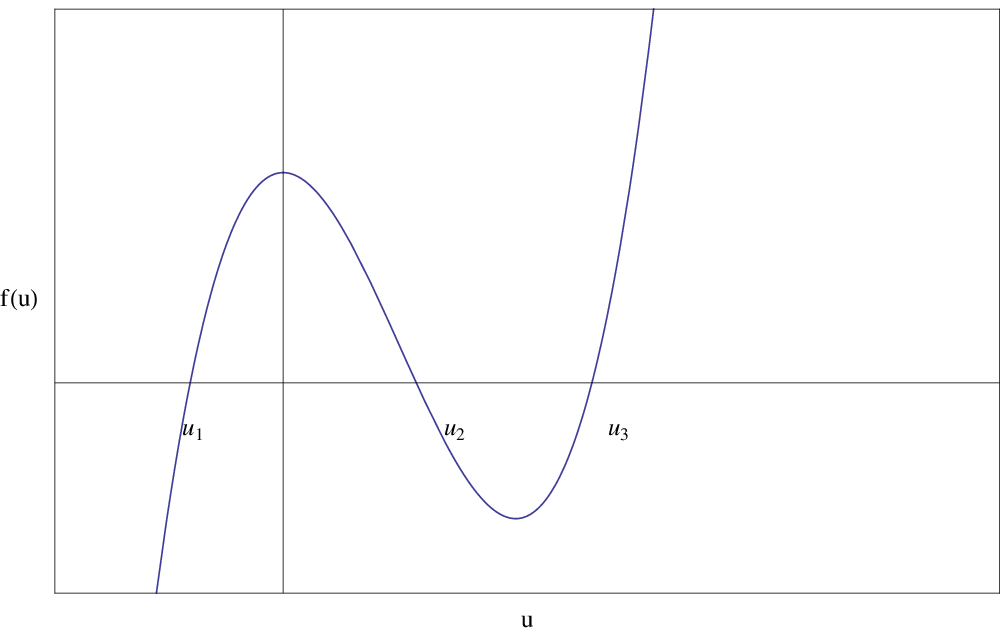}}\\
\vspace{0.2cm}
\end{center}
Figure 5. The graph shows the function $f(u)$ for $ M =0.5, a = 0, L = 30$ and $ E = 9$.\\

In analyzing the  null geodesics of the string black hole, it is clear that the geometry of the geodesics depends on the nature of the roots of the equation $ f(u) =0$. Note that for any value of the parameters in the theory $ M, a, L, E$, the function $f(u) \rightarrow - \infty$ for 
$ u \rightarrow \pm \infty$. Also, for $ u =0$, $f(u) = + \frac{E^2}{L^2}$. Therefore, $f(u)$ definitely has one negative real root ($u_1$). Furthermore, $f(u)$ also has a real root at $ u_4 = 1/a$. This can be observed by the fact that $f(u)$ has a factor  $( -1 + au )$ as follows,
\begin{equation}
f(u) = \frac{ ( -1 + a u)}{L^2} \left( E^2 ( -1 + a u) - L^2 u^2 ( -1 + 2 M u) \right)
\end{equation}
Since one of the roots for $f(u)$ is known, it is possible to write it as
\begin{equation}
f(u) = - 2 M a \left( u - \frac{1}{a} \right) g(u)
\end{equation}
Here, $g(u)$ is a cubic polynomial given by,
$$
g(u) = ( u - u_1) ( u - u_2) ( u - u_3) =$$
\begin{equation}
u^3 - \frac { 1}{ 2 M} u^2- \frac{ a E^2} { 2 M L^2} u + \frac{ E^2}{ 2 M L^2}
\end{equation}
The roots of $g(u)$ are given by $u_1, u_2$ and $u_3$. Since $u_1$ is real, there are two possibilities for $u_2$ and $u_3$: {\it either}, both are real {\it or} they are a complex-conjugate pair. The sum and the products of the roots $u_1, u_2$ and $u_3$ of the polynomial $g(u)$ are related to the coefficients of  $g(u)$ as\cite{math},
\begin{equation}
u_1 + u_2 + u_3 = \frac{1}{ 2 M}
\end{equation}
\begin{equation}
u_1 u_2 u_3 = - \frac{ E^2}{ 2 M L^2}
\end{equation}
As discussed earlier, $u_1$ is real and negative. Therefore, the roots $u_2, u_3$ has to be positive if they are real. This conclusion comes from observing the signs of the eq.(45) and eq.(46). Overall, the polynomial $f(u)$ has a negative($u_1$) and a positive($u_4$) real root always. The other two roots ($ u_2, u_3$) will be either real or complex-conjugate. If they are real, then they will be positive. Also, if they are real, they could be degenerate roots as well.

\subsubsection{ Case 1: Circular orbits}

The conditions for the circular orbits are,
\begin{equation}
\dot{r} =0 \Rightarrow V_{eff} = E_c^2
\end{equation}
and
\begin{equation}
\frac{ d V_{eff}}{dr} =0
\end{equation}
From eq.(48), one get two solutions for circular orbit radius $r$ as,
\begin{equation}
r_{\pm} = \frac{1}{4} \left( a + 6 M \pm \sqrt{ 36 M^2 - 20 a M + a^2} \right)
\end{equation}
$r_+ \geq 2M$ and $r_-  \leq 2M$. This is graphically represented in Fig.6.

\begin{center}
\scalebox{.9}{\includegraphics{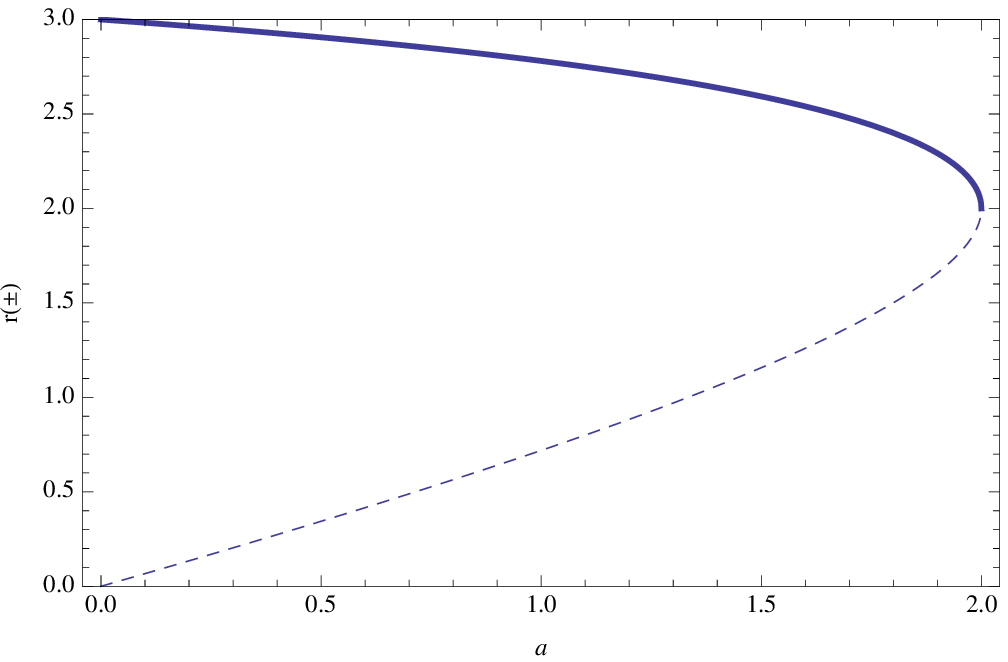}}\\

\vspace{0.2cm}
\end{center}
Figure 6. The graph shows the relation of $r_{\pm}$ with $a$. The dark curve is for $r_+$ and the dashed curve is for $r_-$. Here $ M =1$.\\

Also, $r_{\pm}$ are real when $ a  \leq 2M$ or $ a \geq 18M$. Since we have assumed  $ a < 2 M$ for the purpose of the work in this paper,    radius of the circular orbit that we are interested is  $r_+$ which we will rename  as $r_c$ during the rest of the paper. Hence,
\begin{equation}
r_c  = \frac{1}{4} \left( a + 6 M + \sqrt{ 36 M^2 - 20 a M + a^2} \right)
\end{equation}
The circular orbits at $r=r_c$ are unstable due to the nature of the potential at $ r = r_c$. The hypersurface at $ r = r_c$ is known as the ``photon sphere''. For a detailed discussion about photon spheres see the paper by  Claudel et.al\cite{vir}. When $a \rightarrow 0$, $ r_c \rightarrow 3 M$ which is the radius of the unstable circular orbit of the Schwarzschild black hole\cite{chandra}.

The radius of the circular orbit $r_c$ in eq.(50) is independent of $E$ and $L$. However, they are related to each other from eq.(47) as,
\begin{equation}
\frac{ E_c^2}{ L_c^2} = \frac{f(r_c)} { R^2(r_c)} = \frac{ (r_c - 2 M)} { r_c^2 ( r_c - a)} = \frac{ 1}{ D_c^2}
\end{equation}
Here, $D_c$ is the impact parameter at the critical stage. When $a \rightarrow 0$, $ D_c^2 \rightarrow 27 M^2$ which is the impact parameter for the  unstable circular orbits of the Schwarzschild black hole \cite{chandra}.

One can also study the circular orbits using the function $f(u)$ introduced in section(4.2.2). Circular orbits exists when $f(u)$ has a real degenerate root($u_2 = u_3 = u_c$). Hence, $f(u)$ can be written as,
\begin{equation}
f(u) = - 2 M a ( u - u_c)^2 ( u - \frac{1}{a} ) ( u - u_1)
\end{equation}
Here, $u_c = \frac{1}{r_c}$, where $r_c$ is given by  eq.(50). From eq.(45),
\begin{equation}
u_1 = \frac{1}{2M} - 2 u_c
\end{equation}
The graph for $f(u)$ for this particular case is given in the Fig. 7.

\begin{center}
\scalebox{.9}{\includegraphics{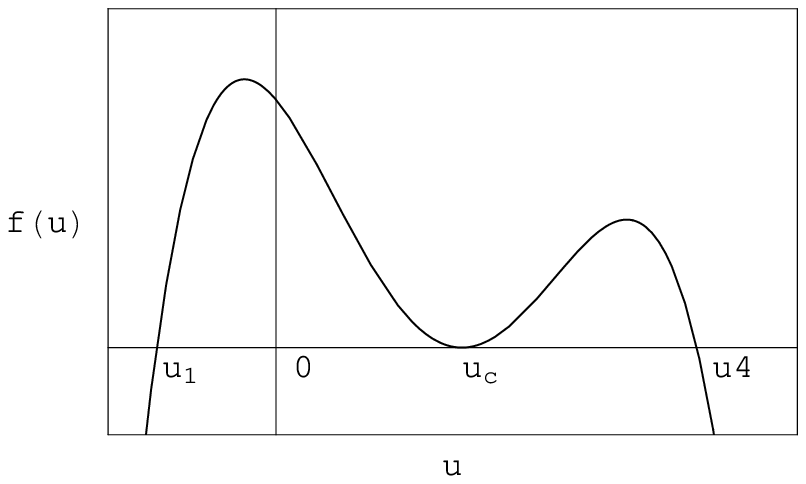}}\\

\vspace{0.2cm}
\end{center}
Figure 7. The graph shows the function $f(u)$ for $ M =0.5, E = 9, a = 0.6, L = 17.7842$. Note that due to the degenerate roots, $E$ and $L$ are related by the eq.(51).\\

For a  null geodesic arriving from infinity and approaching the black hole, the motion is given by the region from $ u \rightarrow 0 ( r \rightarrow \infty)$ to
$ u \rightarrow u_c ( r \rightarrow r_c)$. This is given by the  region  from $u=0$ to $u=u_c$ in Fig. 7. During this region, since $ u \geq 0$ and $ u_1 \leq 0$, $ u - u_1 > 0$. Also, $( u_4 - u) = ( \frac{1}{a} - u) > 0$ for obvious reasons. Therefore, $ f(u) >0$ for $ 0 \leq u \leq u_c$. Hence 
\begin{equation}
\left( \frac{ du}{d \phi} \right) = \pm \sqrt{ f(u) }
\end{equation}
The ``+'' sign will be choosen without lose of generality. One can integrate the equation, $ \frac{ du} { \sqrt{ f(u) }} = d \phi$ to get a relation between $u$ and $\phi$ as,
\begin{equation}
u = \frac{\frac{ u_4}{ c_0^2} tanh^2 \left( \frac{ \phi - \phi_0}{ a_0} \right) + u_1 } { 1 + \frac{1}{c_0^2} tanh^2 \left( \frac{ \phi - \phi_0}{ a_0} \right)}
\end{equation}
Here, $\phi_0$ is a constant of  integration chosen such that when $ u =0$, $\phi =0$. $a_0$ and $c_0$ are constants. $\phi_0$, $a_0$ and $c_0$ are   given by,
\begin{equation}
\phi_0 = - a_0 arctanh \left( \sqrt{ - \frac{u_1}{u_4} } c_0 \right)
\end{equation}
\begin{equation}
a_0 = \sqrt{ \frac{ 2}{ M a} } \frac{ 1}{ \sqrt{ ( u_4 - u_c) ( u_c - u_1) }}
\end{equation}
\begin{equation}
c_0 = \sqrt{ \frac{ u_4 - u_c}{ u_c -u_1} }
\end{equation}
In the Fig. 8, the polar plot of the null geodesics are given for photons arriving from infinity and having an unstable circular orbit at $ r = r_c$.
\begin{center}
\scalebox{.9}{\includegraphics{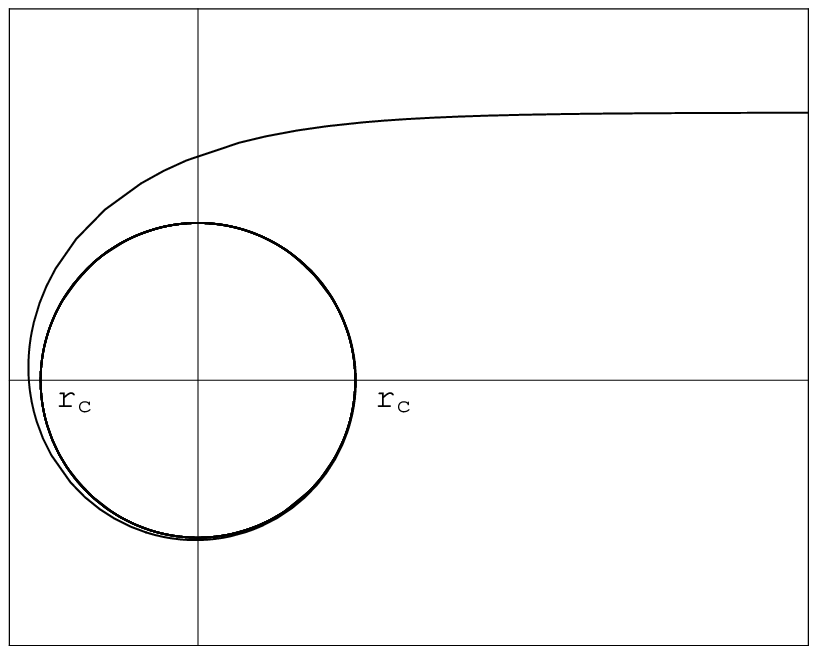}}\\
\vspace{0.2cm}
\end{center}
Figure 8. The polar plot shows the critical null geodesics approaching the blackhole from infinity. The geodesics have an unstable circular orbit at $r =r_c$. Here, $ M = 1, a = 0.2$ and $ r_c = 2.9651$.\\

{\bf The Time Period}\\

The time period for circular orbits can be calculated for proper time as well as coordinate time with $   \phi = 2 \pi$. From eq.(29),
\begin{equation}
T_{\tau} = \frac { 2 \pi r_c ( r_c - a) } { L}
\end{equation}
From combining the two equations in eq.(29), 
\begin{equation}
T_t = \frac{ 2 \pi R(r_c) }{ \sqrt{ f(r_c)}} = \frac{ 2 \pi \sqrt{ r_c^2 ( r_c -a)}} { \sqrt{ r_c  - 2 M}}
\end{equation}
One can compute $T_{\tau}$ and $T_t$  for the Schwarzschild black hole by taking the limit $ a \rightarrow 0$ which results in the following,
\begin{equation}
T_{ t, Sh} = 3 \sqrt{3} M
\end{equation}
and
\begin{equation}
T_{\tau, Sh} = \frac{ 2 \pi ( 3M)^2} { L}
\end{equation}
By observing the graphs of the time periods, it is clear that the periods for the Schwarzschild black hole is larger in comparison with the string black hole.

\begin{center}
\scalebox{.9}{\includegraphics{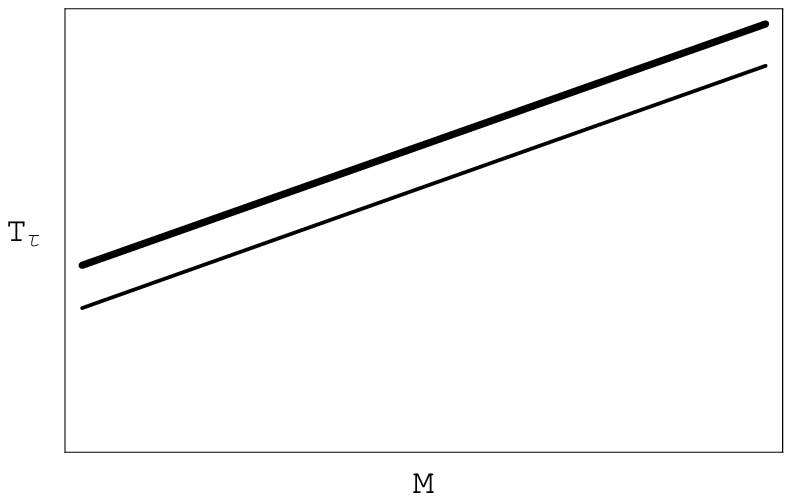}}\\

\vspace{0.2cm}
\end{center}
Figure 9. The graph shows the relation of $T_{\tau}$ with the mass $M$. The dark curve is for $T_{\tau, Sh}$ and the light curve is for $T_{\tau,String}$. Here, $a =1$ and $L =1$.\\

\begin{center}
\scalebox{.9}{\includegraphics{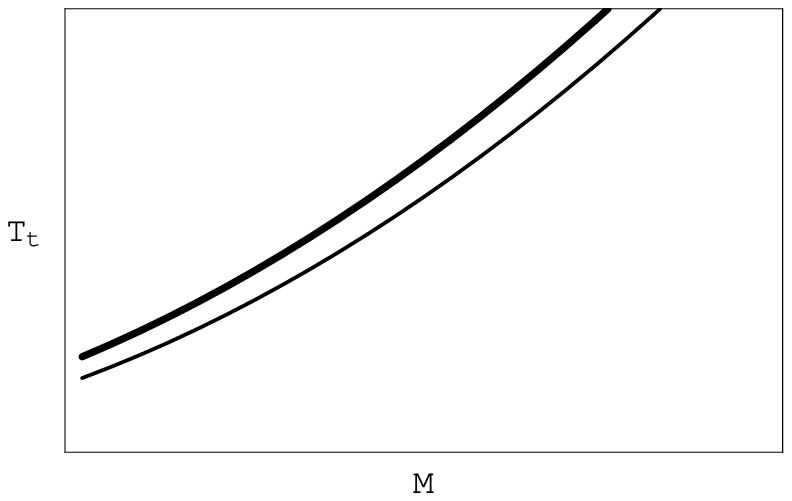}}\\

\vspace{0.2cm}
\end{center}
Figure 10. The graph shows the relation of $T_t$ with $M$. The dark curve is for $T_{t, Sh}$ and the light curve is for $T_{t ,String}$. Here $a =1$ and $L=1$.\\

{\bf The Cone of avoidance}\\

One can define the ``cone of avoidance'' similar to what is described in\cite{chandra} as follows.

The null rays described by the eq.(55), passing a point in the space-time forms the generators of the cone. Let $\Psi$ denote the half-angle of the cone directed towards the black hole at large distances. Then,
\begin{equation}
cot \Psi = + \frac{ d \bar{r} }{ R(r) d \phi}
\end{equation}
Here, $\bar{r}$ is the proper length along the generators of the cone. Hence,
\begin{equation}
d\bar{r} = \frac{1} { \left(1-\frac{2M}{r}\right)} dr
\end{equation}
After substitution, $cot \Psi $ becomes,
\begin{equation}
cot \Psi  = \frac{  1} { \sqrt{(1-\frac{2M}{r}) r ( r -a)}} \frac{ dr}{ d \phi}
\end{equation}
One  can rewrite the above equation in terms of $u = \frac{1}{r}$ and replace 
$\frac{ du}{ d \phi}$ with $ \sqrt{ f(u) }$ given in eq.(54) to obtain,
\begin{equation}
cot \Psi = \frac{ \sqrt{ f(u) } }{\sqrt{ u^2 ( 1 - 2 Mu) ( 1 - au) }}
\end{equation}
Hence the cone of avoidance for the the null geodesics with the critical impact parameter with the solution given in eq.(55) can be written as,
\begin{equation}
tan \Psi =  \frac{\sqrt{ u^2 ( 1 - 2Mu)}} {\sqrt{ 2 M ( u -u_c)^2 ( u -u_1) }}
\end{equation}
From the above equation, it follows that, 
\begin{equation}
for \hspace{1cm} u \rightarrow u_c ( r \rightarrow r_c), \hspace{1 cm} \Psi = \pi/2
\end{equation}
and,
\begin{equation}
 for \hspace{1cm} u \rightarrow 0  ( r \rightarrow \infty), \hspace{1 cm} \Psi  \approx \sqrt{ - \frac{1}{ 2 M u_1 u_c^2 } }\left(\frac{ 1}{r}\right)
\end{equation}
Note that $u_1 < 0$ which makes $\Psi$ real in eq.(69). If we take the limit $ a \rightarrow 0$, the above equation, $\Psi$ approaches $ 3 \sqrt{3} M
\left(\frac{1}{r}\right)$ which is the value for the Schwarzschild black hole given in\cite{chandra}.

\subsubsection{ Case 2 and  3: Unbounded orbits}

${\bf Case\hspace{0.1cm} 2: E = E_2}$

Here, we will study  Case 2 given in the section(4.2.1). In this case, $f(u) =0$ has four real roots. Therefore, the motion is possible in  two regions given as 1 and 2 in the Fig 11.

\newpage
\begin{center}
\scalebox{0.9}{\includegraphics{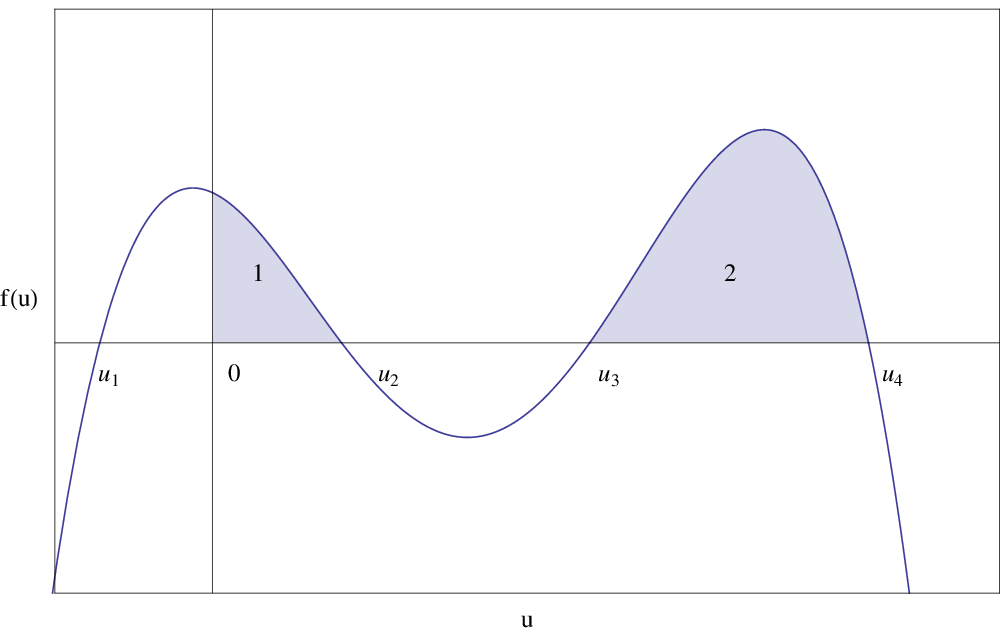}}\\
\vspace{0.2cm}
\end{center}
Figure 11. The graph shows the function $f(u)$ when it has four real roots.\\

If the particle starts at infinity ($r = \infty$ or $ u = 0$), it will fall up to $u_2$ (or $r = r_P$) and fly away to infinity. Since $ \frac{ du} { \sqrt{ f(u) }} = d \phi$, one can integrate both sides to get $\phi$ in terms of Jacobi elliptic integral $\mathcal{F}( \xi, y)$ as,
\begin{equation}
\phi = \frac { - 2 i   \mathcal{F} ( \xi, y)}{ \sqrt{ 2 m a( u_4 - u_1) ( u_3 - u_2)}} + constant
\end{equation}
Here,
\begin{equation}
sin \xi  = \sqrt{ \frac{ ( u - u_2) ( u_4 - u_1) }{ ( u - u_1) ( u_4 - u_2) }}
\end{equation}
\begin{equation}
y = \frac{ ( u_3 - u_1) ( u_4 - u_2)}{ (u_3 - u_2) ( u_4 - u_1)}
\end{equation}
The angle $\phi$ is chosen such that for $ u = u_2$, $ \phi =0$. Hence the integration constant is zero. The corresponding motion is given in Fig 12.

\begin{center}
\scalebox{.9}{\includegraphics{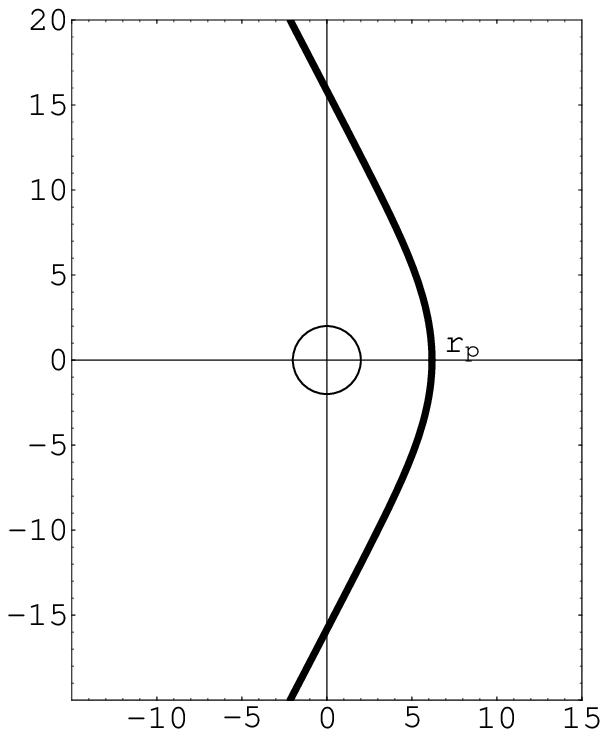}}\\

\vspace{0.2cm}
\end{center}
Figure 12. The polar plot shows the null geodesics approaching the black hole from infinity. The geodesics  Here, $ M =1, a = 0.6, L = 100$ and $ E = 14$.\\

The other unbounded orbit for $E = E_2$ corresponds to the motion starting from $r=r_A$ (or $u=u_3$). Here, the particle will fall into the singularity at $r=a$ (or $u =u_4$) crossing the horizon. In this case, the solutions for  $\phi$ is similar as in eq.(70). Therefore, we will omit the explicit expressions for the $\phi$ in this case.  The integration constant is chosen such that $\phi =0$ when $u=u_3$. In this case, the integration constant is not zero. The corresponding motion is given in Fig 13.

\begin{center}
\scalebox{.9}{\includegraphics{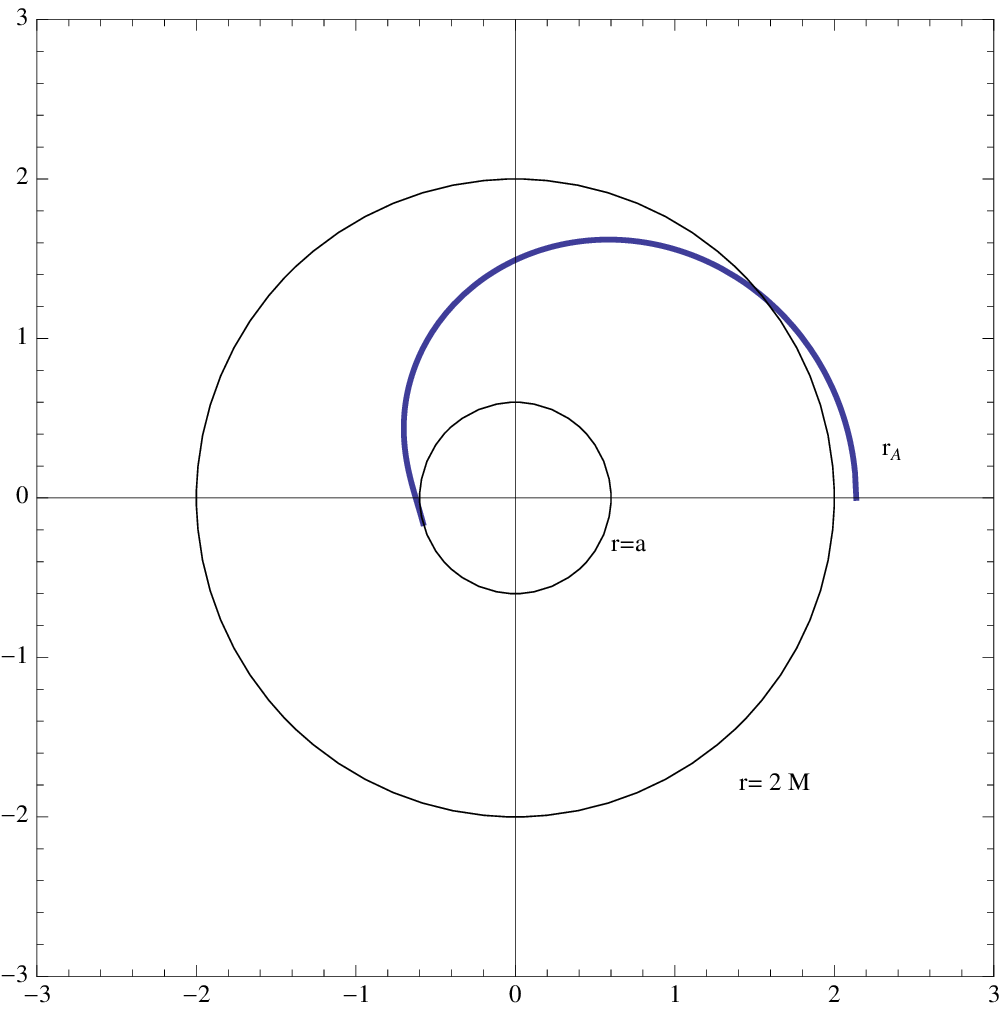}}\\
\vspace{0.2cm}
\end{center}
Figure 13. The polar plot shows the null geodesics falling into  the black hole from $r = r_A$. Here, $ M=1, a=0.6, L=100$ and $E=14$.\\

${\bf Case \hspace{0.1cm} 3: E = E_1}$

Here, we will study the Case 3 given in the section(4.2.1). In this case, $f(u) =0$ has only two real roots as given in Fig 14.  
\begin{center}
\scalebox{.9}{\includegraphics{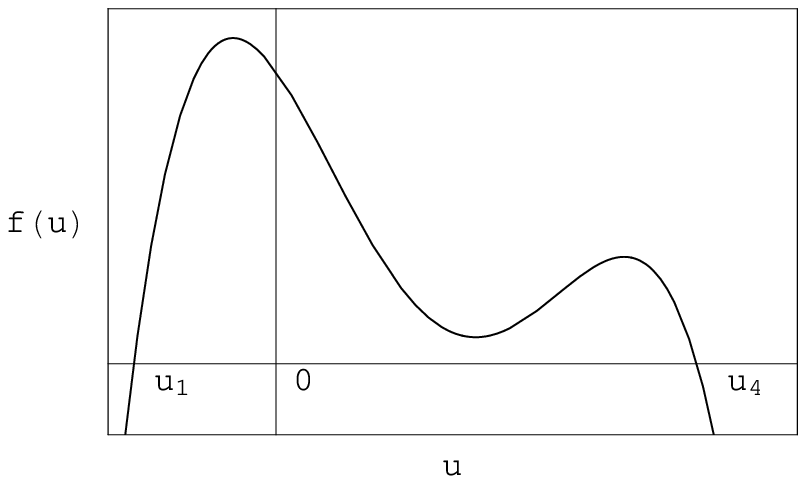}}\\
\vspace{0.2cm}
\end{center}
Figure 14. The graph shows the function $f(u)$ when it has only two real roots.\\

Therefore, the motion is possible in all regions from infinity to the singularity. The integration of  $\frac{ du} { \sqrt{ f(u) }} = d \phi$ gives the same results for $\phi$ as in eq.(70). However, in this case, $u_2$ and $u_3$ are both imaginary and $u_1$ and $u_4$ are real. The integration constant is also imaginary. But, $\phi$ is real. The corresponding motion is given in Fig 15.

\begin{center}
\scalebox{.9}{\includegraphics{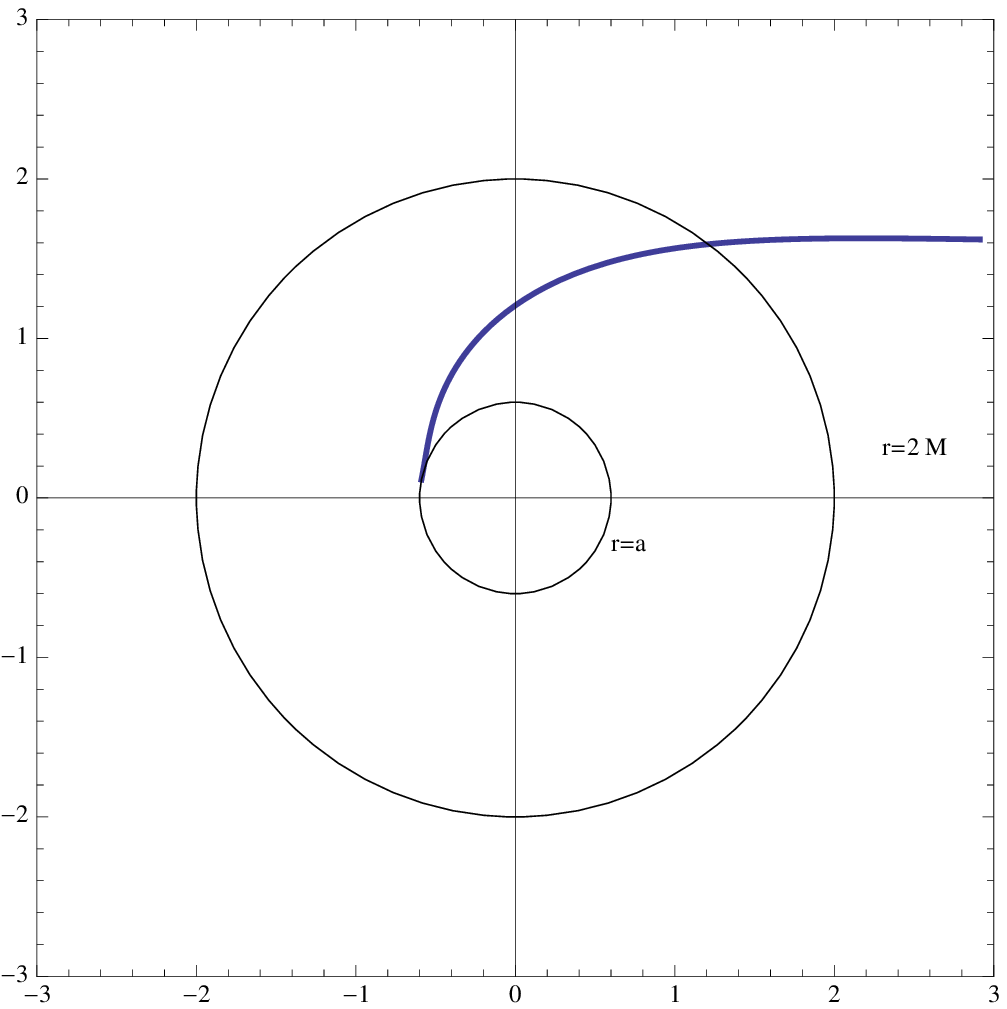}}\\

\vspace{0.2cm}
\end{center}
Figure 15. The polar plot shows the null geodesics approaching the black hole from infinity. The geodesics  Here, $M=1, a=0.6, L=15$ and $E=10$.

\section{ Applications of the null geodesics in the Einstein frame}

In this section we will apply the properties learnt in the previous section to two aspects of null geodesics.

\subsection{ Bending of light by the string black hole}

Motion of photons are represented by the null geodesics and from the knowledge gained in the section 4, one can study important properties related to bending of light. In particular, we will study the  unbounded orbits described in the section 4.2.4 with energy $E = E_2$. The corresponding motion given in Fig.12 displays clearly the bending of light as it travel around the black hole.

First, the closest approach distance $r_o$ is calculated. It is defined by the value of $r$ when $ \frac{dr}{d \phi} =0$. From eq.(40) and eq.(43),
\begin{equation}
\left(\frac{1}{r^2} \frac{dr}{d \phi} \right)^2 = f(r) = \frac{ 2 M}{r} ( r - a) g(r)
\end{equation}
Since $r=a$ is inside the black hole, the roots of $\frac{dr}{d \phi} =0$ which yields $r_o$ corresponds to a root of $g(r)=0$ polynomial which is given by,
\begin{equation}
r^3  - a r^2 - D^2 r + 2 M D^2 =0
\end{equation}
From Fig.11, one can conclude that the roots of the above equation corresponds to $ r_1, r_2, r_3$ which are the inverse of $ u_1, u_2, u_3$. Since $r_1 <0$, the roots to be considered are $r_2$ or $r_3$. Note that $ r_3 < r_2$ (since $u_3 > u_2$). Both $r_2, r_3$ are greater than the horizon radius. However the radius of the unstable circular orbit $r_c > r_3$. Hence, the root we will choose for the closest approach is $r_2$. Note that this argument is clear if one study Fig.3 as well. Here the closest approach $r_o$ corresponds to $r_P$, not $r_A$.

Now, one can use the well known techniques in determining the roots of a cubic polynomial to obtain $r_0$ as,
\begin{equation}
r_o^{string} = 2 \sqrt{- \frac{ p}{3} } cos \left( \frac{1}{3} cos^{-1} \left( \frac{ 3 q}{2 p} \sqrt{ -\frac{ 3}{p}} \right) \right) + \frac{a}{3}
\end{equation}
Here, $p$ and $q$ are given by,
\begin{equation}
 p = \frac{ a^2 - 3 D^2}{3}
\end{equation}
\begin{equation}
q = \frac{ 54 M D^2 - 9 a D^2 - 2 a^3 }{ 27}
\end{equation}
When $ a \rightarrow 0$, $r_o$ approaches the well known value for the Schwarzschild black hole given by \cite{tolu},
\begin{equation}
r_o^{Sch} =  \frac{ 2 D}{ \sqrt{3}}  cos \left( \frac{1}{3} cos^{-1} \left( \frac{ -\sqrt{27} M}{D} \right) \right)
\end{equation}

\begin{center}
\scalebox{.9}{\includegraphics{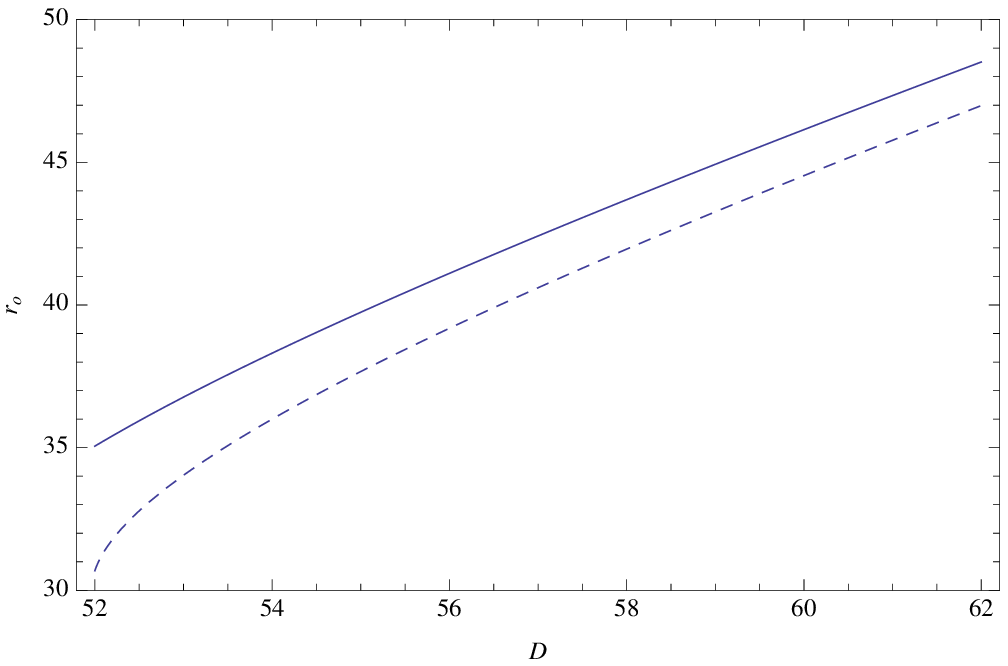}}\\
\vspace{0.2cm}
\end{center}
Figure 16. The graph shows the closest approach $r_o$ for the string black hole(dark curve)  and the Schwarzschild black hole(dashed curve)  as a function of the impact parameter $D$. Here, $M = 10$  and $a = 2$.\\

Now, we will compute the  angle of deflection  of light for the GMGHS black hole. In a paper by Bhadra \cite{bhadra}, the deflection angle was given as,
$$
\alpha_{string} = \frac{ 4 M}{ r_{o}} + 
\frac{ 4 M^2}{ r_{o}}\left( \frac{ 15 \pi}{16} -1 \right) +\frac{ a M}{r_{o}^2} \left( \frac{ 3 \pi}{4} -2 \right)
$$
\begin{equation}
+ \frac{ a^2}{r_{o}^2} \left( 2 - \frac{\pi}{16} \right)
\end{equation}
Note that here, $r_o$ is the one given in eq.(75) for the string black hole.

When $ a \rightarrow 0$, one obtain the well known bending angle for the  Schwarzschild black hole, as,
\begin{equation}
\alpha_{Sch} = \frac{ 4 M}{ r_o} + 
\frac{ 4 M^2}{ r_o}\left( \frac{ 15 \pi}{16} -1 \right) 
\end{equation}
Here, $r_o$ corresponds to the one given in eq.(78).
In the paper by Bhadra \cite{bhadra}, the deflection angle was given as a function of the closest approach. Here, we will substitute the expression for respective expressions for $r_0$ for both black holes and plot the deflection angle as a function of the impact parameter $D$. 
\begin{center}
\scalebox{.9}{\includegraphics{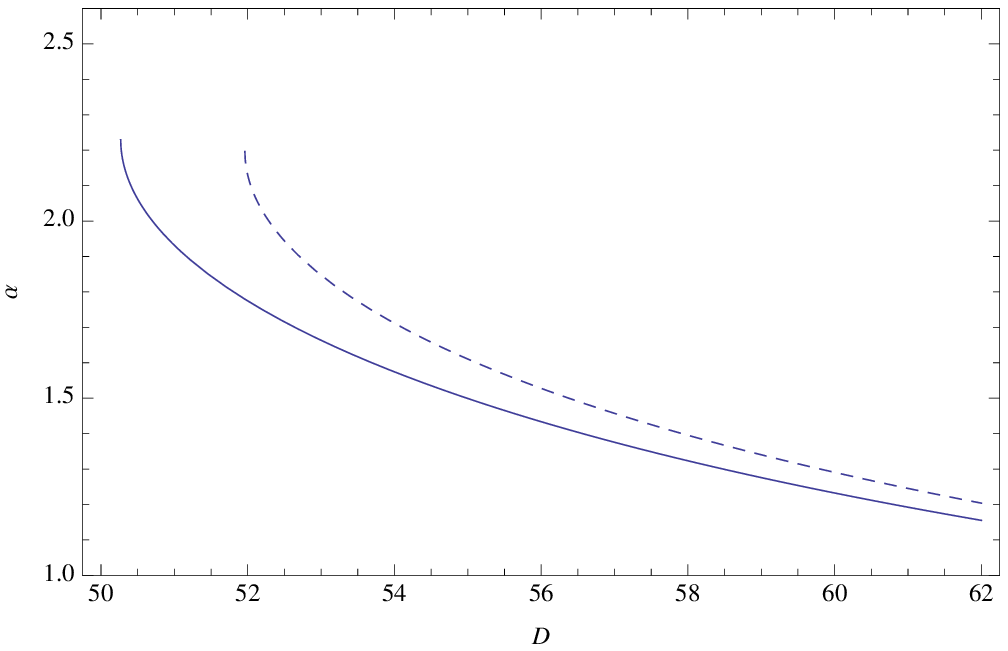}}\\
\vspace{0.2cm}
\end{center}
Figure 17. The graph shows the bending angle $\alpha$ for the string black hole(dark curve)  and the Schwarzschild black hole(dashed curve)  as a function of the impact parameter $D$. Here, $M = 10$  and $a = 2$.\\

From the above Fig. 17, it is clear that the photons with the same impact parameter bends less around the string black hole compared to the Schwarzschild black hole.

\subsection{Unstable null geodesics and quasinormal modes of massless scalar field in the eikonal limit}

When a black hole is perturbed, it undergoes damped oscillations and the frequencies of oscillations are called quasinormal modes. Due to a wide variety of applications, there are many works that are published in computing the quasinormall modes of various black holes. A good review is given by Konoplya \cite{kon}. In this section we will present the quasinormal mode frequencies of the string black hole for a massless scalar field following the paper by Cardoso et.al \cite{car}.

First, let us give some back ground information. The  equation for a massless scalar field in the back ground of the GMGHS black hole is given by,
\begin{equation}
\frac{ d^2 \eta}{dr_*^2} + Q_o \eta =0
\end{equation}
Here, 
\begin{equation}
Q_o = \omega^2 - V_{scalar}(r)
\end{equation}
where,
\begin{equation}
V_{scalar}(r) = \frac{ l(l +1)}{R^2} + \frac{ f f' R'}{R } + \frac{ f^2 R''}{R}
\end{equation}
Here, $l$ is the spherical harmonic index and $r_*$ is the tortoise coordinate. More information on the massless scalar perturbation of the GMGHS black hole can be found in \cite{fer2}.
In the eikonal limit($ l \rightarrow \infty$), 
\begin{equation}
Q_o \approx \omega^2 - \frac{ f l}{R^2}
\end{equation}
Observing eq.(84), one can conclude that the maximum of $Q_o$ occurs at $r = r_m$ given by,
\begin{equation}
2 f(r_m) R'(r_m) - R(r_m) f(r_m) =0
\end{equation}
However, since the effective potential for the null geodesics, is given by $V_{eff} = \frac{L^2 f}{R^2}$, the unstable circular orbits occurs at $V_{eff}'=0$ leading to,
\begin{equation}
2 f(r_c) R'(r_c) - R(r_c) f'(r_c) =0
\end{equation}
Hence the maximum of $Q_o$ and the location of the null circular geodesics coincides at $ r_m = r_c$. Therefore, the computation of quasinormal modes at the eikonal limit and the unstable null geodesics are related. Cardoso et.al.\cite{car} presented an important result based on this: In the eikonal limit, the quasinormal modes are given as,
\begin{equation}
\omega_{QNM} = \Omega_c l - i ( n + \frac{1}{2} ) | \lambda|
\end{equation}
Here, $\Omega_c$ is the coordinate angular velocity given by $ \frac{\dot{\phi}}{\dot{t}}$ computed at $ r = r_c$. $\lambda$ is the Lyapunov exponent which gives the instability timescale of the unstable circular null geodesics. We will  not present the derivation of the above results since it is clearly done in Cardoso et.al\cite{car}. For the GMGHS black hole, $\Omega_c$ and $\lambda$ are given as,
\begin{equation}
\Omega_c = \frac{ \dot{\phi(r_c)}}{\dot{t}(r_c)}  = \sqrt{ \frac{ f(r_c)}{R(r_c)^2 }} 
= \sqrt{ \frac{ r_c - 2M}{r_c^2(r_c -a)}}
\end{equation}

$$
\lambda = \sqrt{ \frac{ -V_{eff}''(r_c)}{ 2 \dot{t}(r_c)^2}} 
= \sqrt{ \frac{-V_{eff}''(r_c) R(r_c)^2 f(r_c)}{2 L^2} }
$$
\begin{equation}
= \sqrt{ \frac{ ( 2 M -r_c) ( 3 r_c^3 - 12 Mr_c^2 - 3 a r_c^2 + 16 a M r_c + a^2 r_c - 6 a^2 M)} { r_c^4 ( r_c -a)^2 } }
\end{equation}

\begin{center}
\scalebox{.9}{\includegraphics{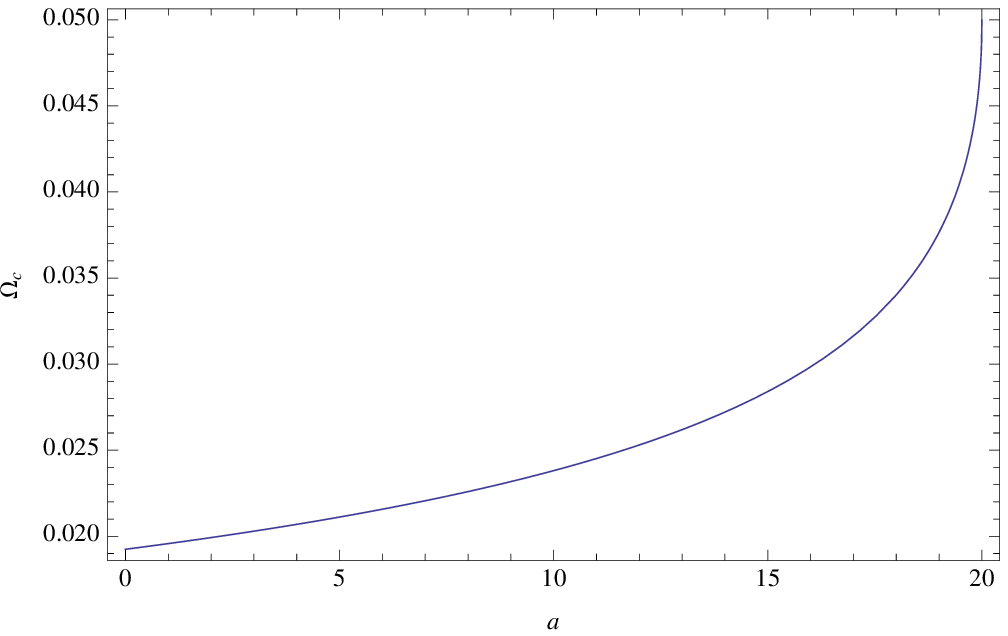}}\\
\vspace{0.2cm}
\end{center}
Figure 18. The graph shows $\Omega_c$ as a function of $a$. Here, $M = 10$\\

\begin{center}
\scalebox{.9}{\includegraphics{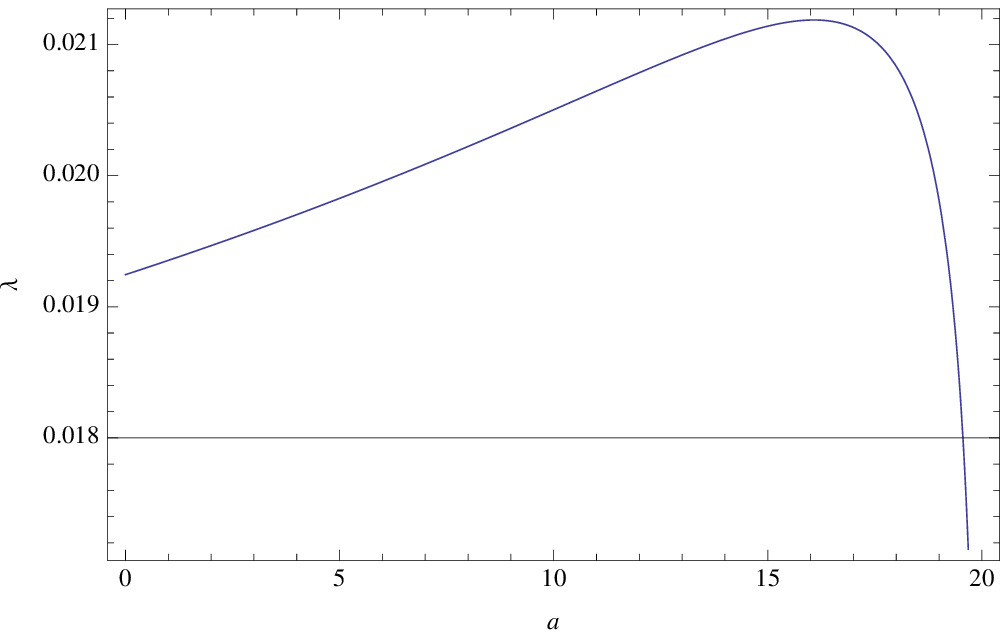}}\\
\vspace{0.2cm}
\end{center}
Figure 19. The graph shows the Lyapunov exponent $\lambda$ as a function of $a$.
Here, $M = 10$.\\

Having computed $\Omega_c$ and $ \lambda$, one can extract the real and the imaginary part of $\omega$ from eq.(87) easily. Note that $\lambda$ has a maximum at $ a = 6 M ( 2 - \sqrt{3})$.

\section{ Geodesics in String Frame}

The approach in deriving the geodesics for the string frame is similar to the
approach for the geodesics in the Einstein frame. First, let us rewrite the metric in the string frame as,
\begin{equation}
ds^2 = -f(r) dt^2 + g(r)^{-1} dr^2 +  r^2 ( d \theta^2 + sin^2(\theta) d \phi^2)
\end{equation}
Here,
\begin{equation}
f(r) = \frac{\left(1-\frac{2m}{r}\right)}{ \left( 1 + \frac{ 2 m \sinh^2(\alpha)}{r} \right)^2} =\frac{\left(1-\frac{2m}{r}\right)}{ \left( 1 + \frac{ 2 b}{r} \right)^2}
\end{equation}
\begin{equation}
g(r) = \left(1-\frac{2m}{r}\right)
\end{equation}
\begin{equation}
b = m \sinh^2(\alpha)
\end{equation}
Note that we will use $r$ instead of $\hat{r}$ here on. Equations governing the geodesics in this space-time can be derived from the Lagrangian equation,
\begin{equation}
{\cal{L}} =  - \frac{1}{2} \left( - f(r) \left( \frac{dt}{d\tau} \right)^2 +  \frac{1}{g(r)}\left( \frac{dr}{d \tau} \right)^2 + r^2 \left(\frac{d \theta}{d \tau} \right)^2 + r^2 sin^2 \theta \left( \frac{d \phi }{d \tau} \right)^2 \right)
\end{equation}
Here, $\tau$ is an affine parameter along the geodesics. The canonical momenta corresponding to each coordinate is,
\begin{equation}
p_t = \frac{d {\cal{L}} }{d \dot{t}} = f \dot{t}
\end{equation}
\begin{equation}
p_{r} = - \frac{d {\cal{L}} }{d \dot{r}} = \frac{ \dot{r}}{g}
\end{equation}
\begin{equation}
p_{\theta}  = - \frac{d {\cal{L}} }{d \dot{\theta}} =  r^2 \dot{\theta}
\end{equation}
\begin{equation}
p_{\phi}  = - \frac{d {\cal{L}} }{d \dot{\phi}} =  r^2 sin^2 \theta \dot{\phi}
\end{equation}
Since the GMGHS black hole have  two Killing vectors $\partial_t$ and $\partial_{\phi}$, there are two constants of motion which can be labeled as $E$ and $L$ given as,
\begin{equation}
p_t = f \dot{t} = E
\end{equation}
\begin{equation}
p_{\phi} =    r^2 sin^2 \theta \dot{\phi} = L
\end{equation}
Furthermore, from the Euler-Lagrangian  equation of motion,
\begin{equation}
\frac{d}{d \tau} \left(\frac{d {\cal{L}} }{d \dot{\theta}} \right) - \frac{d {\cal{L}} }{d \theta}=0 \Rightarrow -\frac{d p_{\theta}}{d \tau}  - \frac{d {\cal{L}} }{d \theta}=0
\Rightarrow 
-\frac{d (r^2 \dot{\theta}) }{d \tau} + r^2 sin \theta  cos \theta  \left( \frac{d \phi }{d \tau}\right)^2=0
\end{equation}
Hence, if we choose $\theta = \pi/2$ and $\dot{\theta} = 0$ as the initial conditions,  the eq.(82) leads to,
\begin{equation}
r^2 \ddot{\theta} +  \frac{d r^2 }{d \tau} \dot{ \theta} = 0
\end{equation}
Therefore, $\ddot{\theta} =0$. $\theta$ will remain at $\pi/2$ and the geodesics will be described in an invariant plane at $ \theta = \pi/2$.
From eq.(99) and eq.(100),
\begin{equation}
r^2 \dot{\phi} = L; \hspace{1.0cm} f(r) \dot{t} = E
\end{equation}
With $\dot{t}$ and $\dot{\phi}$ given by equation(101), the Lagrangian in eq.(94) simplifies to be,
\begin{equation}
\dot{r}^2 + g(r) \left(  \frac{L^2}{r^2} + h   \right)  = E^2 \frac{g(r)} {f(r) }
\end{equation}
We have replaced $2{\cal{L}}=h$. $h=1$  corresponds to time-like geodesics and $h=0$ corresponds to null geodesics. For a time-like geodesics, $\tau$ may be identified with proper time of the particle describing the geodesics. Comparing eq.(85) with $\dot{r}^2 + V_{eff}= 0$, one get the effective potential, which depend on $E$ and $L$ as follows:
\begin{equation}
V_{eff} = \left(  \frac{L^2}{r^2} + h  \right) g(r) - E^2 \frac{ g(r) } {f(r)}
\end{equation}
By eliminating the parameter $\tau$ from the equations (103) and (105), one can get a relation between $\phi$ and $r$ as follows;
\begin{equation}
\frac{ d \phi} {d r} =  \frac{L}{r^2} \frac{1}{ \sqrt{  - V_{eff}} }
\end{equation}

\section{Null Geodesics in String frame}

In the above formalism, $h =0$ for null geodesics.

\subsection{ Radial null geodesics}

The radial geodesics corresponds to the motion of particles with zero angular momentume $(L =0)$. Hence for radial null geodesics, the effective potential,
\begin{equation}
V_{eff} =  - E^2 \frac{g(r)}{f(r)}
\end{equation}
The two equations for $\dot{t}$ and $\dot{r}$ simplifies to,
\begin{equation}
\dot{ r} = \pm \sqrt{ \frac{g}{f}} E = \pm E \left( 1 + \frac{ 2 b}{r}\right)
\end{equation}
\begin{equation}
 \dot{t} = \frac{E}{f(r)} = E \frac{ ( 1 + \frac{ 2 b}{r} )^2}{ (1 - \frac{ 2 m}{r})}
\end{equation}
The above two equations lead to,
\begin{equation}
\frac{ dt}{ dr} = \pm\frac{1}{\sqrt{g(r) f(r)} } = \pm \frac{ (1 + \frac{2 b} {r})} { (1 - \frac{ 2 m}{ r}) }
\end{equation}
The above equation can be integrated to give the coordinate time $t$ as a function of $r$ as,
\begin{equation}
t = \pm \left( r + (2 m + 2 b) ln \left( \frac{r}{ 2m} -1 \right) \right) + const_{\pm}
\end{equation}
When $ r \rightarrow 2 m$, $ t \rightarrow \infty$. On the other hand, the proper time can be obtained by integrating,
\begin{equation}
\frac{ d \tau}{ dr} = \pm\frac{1}{ E ( 1 + \frac{ 2 b}{r})} 
\end{equation}
which leads to,
\begin{equation}
 \tau = \pm \frac{1}{E} \left(r -  2b ln( 2b +r) \right) + const_{\pm}
\end{equation}
When $ r \rightarrow 2 m$, $ \tau \rightarrow \pm  (2 m -  2b ln( 2b + 2m ))/E$, which is finite. Hence the proper time is finite while the coordinate time is infinite. This is the similar to what was   obtained in the Einstein frame.

\subsection{ Geodesics with angular momentum ( $ L \neq 0$ )}

In this section, we will study the null geodesics with angular momentum.

\subsubsection{ Effective potential}

In this case, 
\begin{equation}
 V_{eff} = g(r) \frac{L^2}{ r^2} - E^2 \frac{ g(r)}{f(r)}
\end{equation}
In the Fig. 20, the  $V_{eff}$ is given for various values of $b$. One can note, that  the height is lower for the string black hole in comparison to  the Schwarzschild black hole.

\begin{center}
\scalebox{.9}{\includegraphics{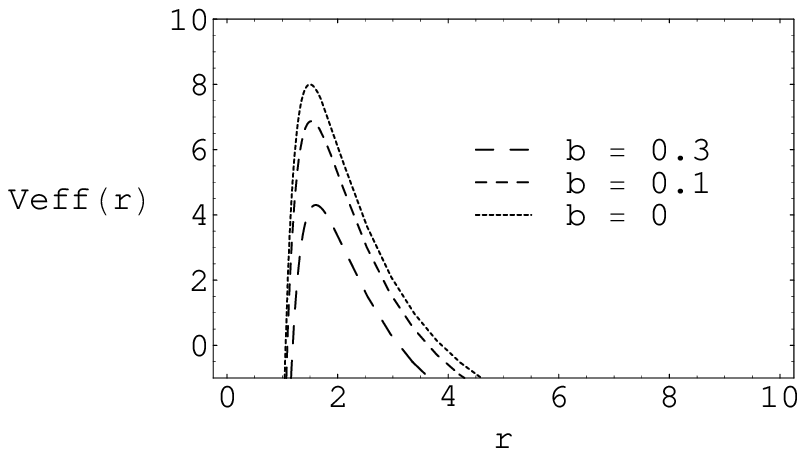}}\\
\vspace{0.2cm}
\end{center}
Figure 20. The graph shows the relation of $V_{eff}$ with the parameter $b$. Here, $m =0.5, L =12$ and $E = 2$.\\

\subsubsection{ Circular orbits}

The conditions for the circular orbits are,
\begin{equation}
\dot{r} =0 \Rightarrow V_{eff} = 0
\end{equation}
and
\begin{equation}
\frac{ d V_{eff}}{dr} =0
\end{equation}
From eq.(116), one get a  solution for circular orbit radius $r$ as,
\begin{equation}
r_{c} = \frac{1}{2} \left( 3 m +  \sqrt{ 8 b m + 9 m^2} \right)
\end{equation}
The circular orbits at $r=r_c$ are unstable due to the nature of the effective potential at $ r = r_c$. The radius of the circular orbit is independent of $E$ and $L$. However, they are related to each other from eq.(115) as,
\begin{equation}
\frac{ E_c^2}{ L_c^2} = \frac{f(r_c)} { r_c^2} = \frac{ (r_c - 3 m)} { 2 b r_c ( r_c + 2 b)}
\end{equation}
When $b \rightarrow 0$, $ r_c \rightarrow 3 m$ which is the radius of the unstable circular orbit of the Schwarzschild black hole\cite{chandra}.

\newpage

\begin{center}
\scalebox{.9}{\includegraphics{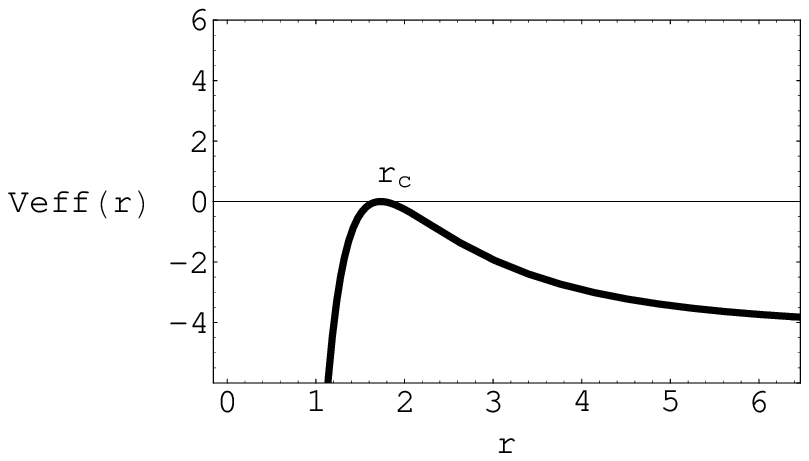}}\\
\vspace{0.2cm}
\end{center}
Figure 21. The graph shows  $V_{eff}$. Here, $m =0.5, b =0.4$ and $E = 2$. The corresponding $L = 7.79$ and $r_c = 1.73$.\\

One can also study the orbits by doing a well known change of variable $ u = \frac{1}{r}$. Then the eq.(106) can be rewritten in terms of $u$ and $\phi$ as,
\begin{equation}
\left(\frac{ d u}{ d \phi} \right)^2 = f(u)
\end{equation}
where,
\begin{equation}
f(u) = 2  m u^3  + u^2 \left( -1 + 4 b^2 \frac{ E^2}{L^2} \right) + 4 b \frac{E^2}{L^2} u + \frac{ E^2}{L^2}
\end{equation}
In analyzing the string black hole null geodesics, it is clear that the geometry of the geodesics depends on the nature of the roots of the equation $ f(u) =0$. Note that for any value of the parameters $ m, b, L, E$, the function $f(u) \rightarrow  \infty$ for 
$ u \rightarrow + \infty$ and $f(u) \rightarrow  - \infty$ for 
$ u \rightarrow - \infty$. Also, for $ u =0$, $f(u) = + \frac{E^2}{L^2}$. Therefore, $f(u)$ definitely has one negative real root ($u_1$). Since $f(u)$ is a cubic polynomial, one can rewrite it as,
\begin{equation}
f(u) = 2 m ( u - u_1) ( u - u_2) ( u - u_3)
\end{equation}
The roots of $f(u)$ are given by $u_1, u_2$ and $u_3$. Since $u_1$ is real, there are two possibilities for $u_2$ and $u_3$: {\it either}, both are real {\it or} they are a complex-conjugate pair. The sum and the products of the roots $u_1, u_2$ and $u_3$ of the polynomial $f(u)$ are related to the coefficients of  $g(u)$ as \cite{math},
\begin{equation}
u_1 + u_2 + u_3 = \frac{1}{ 2 m} - \frac{ 2 b^2}{m} \frac{ E^2}{L^2}
\end{equation}
\begin{equation}
u_1 u_2 u_3 = - \frac{ E^2}{ 2 m L^2}
\end{equation}
As discussed earlier, $u_1$ is real and negative. Therefore, both roots $u_2, u_3$ will be either positive or  negative if they are real. This conclusion comes from observing the sign of the eq.(123). If they are real, then they could be degenerate roots as well as given in Fig. 20. The function $f(u)$ for general values of $ m, b, E, L$ is given in the Fig.22 and 23. 

\begin{center}
\scalebox{.9}{\includegraphics{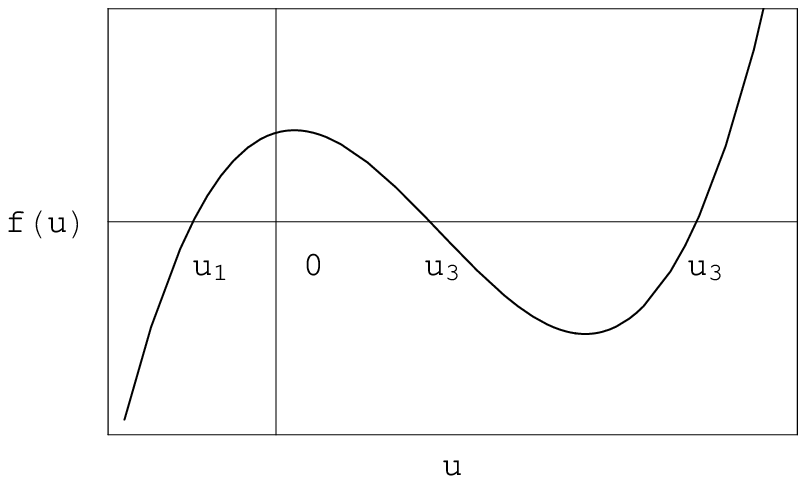}}\\

\vspace{0.2cm}
\end{center}
Figure 22. The graph shows the function $f(u)$ for $ m =0.5, b = 0.4, E = 2$ and $ L = 9.8$.\\

\begin{center}
\scalebox{.9}{\includegraphics{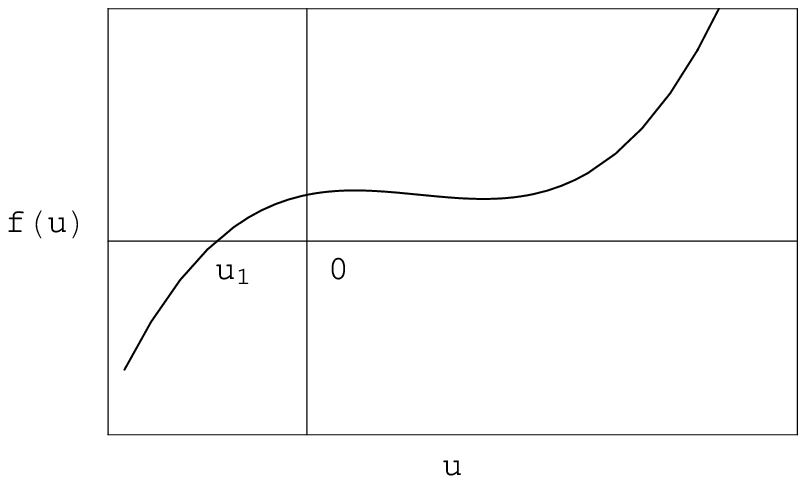}}\\

\vspace{0.2cm}
\end{center}
Figure 23. The graph shows the function $f(u)$ for $ m =0.5, b = 0.4, E = 2$ and $ L = 5.7$.\\

When $ b \rightarrow 0$, $ f(u) \rightarrow 2 mu^3 - u^2 + \frac{E^2}{L^2}$ as expected for the Schwarzschild black hole\cite{chandra}.

Circular orbits exists when $f(u)$ has  real degenerate roots ($u_2=u_3=u_c$). The graph for this particular case is given in the Fig.24.

\begin{center}
\scalebox{.9}{\includegraphics{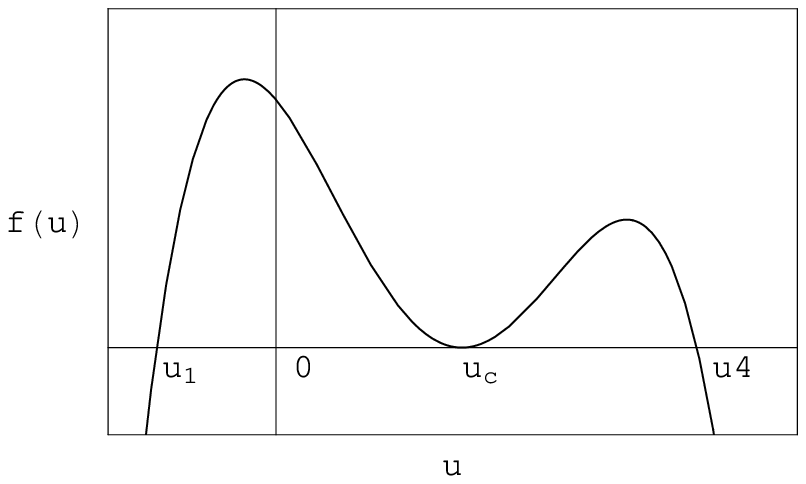}}\\
\vspace{0.2cm}
\end{center}
Figure 24. The graph shows the function $f(u)$ for $ m =0.5,  b = 0.4, E = 2$ and $ L =7.79$. $u_c = 0.578$.  Note that due to the degenerate roots, $E$ and $L$ are related by the eq.(108).\\

Hence, $f(u)$ can be written as,
\begin{equation}
f(u) =  2 m ( u - u_c)^2  ( u - u_1)
\end{equation}
Here, $u_c = \frac{1}{r_c}$, where $r_c$ is given by  eq.(117). From eq.(122),
\begin{equation}
u_1 = \frac{1}{2m} - 2 u_c - \frac{ 2 b^2}{m} \frac{ E_c^2}{L_c^2}
\end{equation}
For a  null geodesic arriving from infinity and approaching the black hole, the motion is given by the region from $ u \rightarrow 0 ( r \rightarrow \infty)$ to
$ u \rightarrow u_c ( r \rightarrow r_c)$. This is given by the  region  from $u =0$ to $u =u_c$ in Fig.24. During this region, since $ u \geq 0$ and $ u_1 \leq 0$, $ u - u_1 > 0$. Therefore, $ f(u) >0$ for $ 0 \leq u \leq u_c$. Hence 
\begin{equation}
\left( \frac{ du}{d \phi} \right) = \pm \sqrt{ f(u) }
\end{equation}
The ``+'' sign will be choosen without lose of generality. One can integrate the equation, $ \frac{ du} { \sqrt{ f(u) }} = d \phi$ to get a relation between $u$ and $\phi$ as,
\begin{equation}
u = u_1 + ( u_c - u_1)  tanh^2 \left( \frac{ \phi - \phi_0}{ a_0} \right) 
\end{equation}
Here, $\phi_0$ is a constant of  integration chosen such that when $ u =0$, $\phi =0$. $a_0$ is also a constants. $\phi_0$ and $a_0$   are  given by,
\begin{equation}
\phi_0 = a_0 arctanh \left( \sqrt{  \frac{u_1}{u_1 - u_c} }  \right)
\end{equation}
\begin{equation}
a_0 = -\sqrt{ \frac{ 2}{ m  ( u_c - u_1) }}
\end{equation}
In the Fig.25, the polar plot of the null geodesics are given for photons arriving from infinity and having an unstable circular orbit at $ r = r_c$.
\newpage

\begin{center}
\scalebox{.9}{\includegraphics{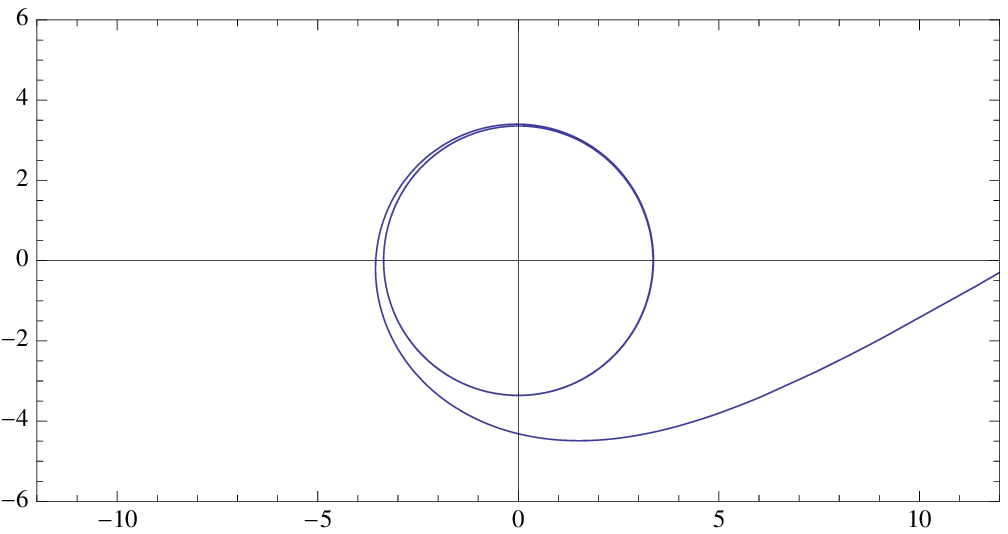}}\\

\vspace{0.4cm}
\end{center}

Figure 25. The polar plot shows the critical null geodesics in the string frame approaching the black hole from infinity. The geodesics have an unstable circular orbit at $r=r_c$. Here, $m=1, b=0.2$ and $r_c=3.1279$.\\

\section{ Conclusions}

We have studied the null geodesics of the static charged black hole in heterotic string theory. The equations for the geodesics were solved exactly for various values of energy and angular momentum of the photons. In the Einstein frame, all possible motions were presented. The circular orbits were shown to be unstable.
In comparison, the circular orbits of the Reissner-Nordstr$\ddot{o}$m black hole which is the charged black hole in general relativity is also shown to be unstable as worked out  by Chandrasekhar\cite{chandra}. The circular orbit of the geodesics in the string frame is also studied in detail in this paper. Since the null geodesics do not change under conformal transformations, the  circular orbits in both frames are expected to be similar which is shown analytically in this paper.

As physical applications of the properties of the null geodesics obtained here, we have studied light bending and quasinormal  modes of massless scalar filed. The closest approach of the photons bending around the black hole is computed as a function of the impact parameter. The deflection angle $\alpha$ is computed as a function of the impact parameter. A comparison is done with the deflection angle of the Schwarzschild black hole. It was observed  that the photons with the same impact parameter bends less around the string black hole compared to the Schwarzschild black hole. These results would be beneficial in computations of gravitational lensing of string black holes. 

The unstable circular null geodesics of the black hole is used to compute the quasinormal modes of the black hole in the eikonal limit. We have followed an important result by Cardoso et.al.\cite{car} in deriving these results. The Lyapunov exponent $\lambda$, which gives the instability time scale is also computed. It was noted that there is a maximum value for $\lambda$ at $ a = 6 M ( 2 - \sqrt{3})$.

As an extension of this work, it would be  interesting to   study the motion for the naked singularity of the charged string solutions considered in this paper. The motion of test particles around the naked singularity of the Reissner-Nordstr$\ddot{o}$m space-times have been studied by Pugliese et.al.\cite{pug1}.

\vspace{0.3cm}

{\bf Acknowledgments}: This work was done during the sabbatical leave granted to the author by Northern Kentucky University. The author likes to  thank Dr. Don Krug of the Mathematics Department of Northern Kentucky University for valuable discussions at the initial stage of this work.

\end{document}